\documentclass[12pt,a4paper]{article}
\pdfoutput=1
\usepackage{latexsym, amssymb, amscd, amsthm, amsxtra, amsmath,amsthm,mathtools }
\usepackage[latin1]{inputenc}
\usepackage{amsfonts}
\usepackage{graphicx}
\usepackage{natbib}
\usepackage[left=0.75in, right=0.75in, top=0.75in, bottom=0.75in]{geometry}
\usepackage{setspace}
\usepackage{algorithm}
\usepackage{algpseudocode}
\usepackage{pifont}
\usepackage{url}
\usepackage{multirow}
\usepackage{epstopdf}

\theoremstyle{plain}
\newtheorem{thm}{Theorem}[section]

\def\beq{\begin{equation}}
\def\eeq{\end{equation}}
\def\bs{\begin{eqnarray*}}
	\def\es{\end{eqnarray*}}
\def\beqr{\begin{eqnarray}}
\def\eeqr{\end{eqnarray}}

\def\bu{{\bf u}}
\def\vv{{\bf v}}
\def\bv{{\bf v}}

\def\bx{{\bf x}}

\def\by{{\bf y}}

\def\bell{\boldsymbol \ell}
\def\bone{\mathbf 1}
\def\bzero{\mathbf 0}

\def\bsX{\boldsymbol{X}}
\def\bsY{\boldsymbol{Y}}
\def\bX{\mathbf X}
\def\bS{\mathbf S}
\def\bY{\mathbf Y}
\def\bA{\mathbf A}
\def\bI{\mathbf I}
\def\bJ{\mathbf J}

\def\bZ{\mathbf Z}

\def\bB{\mathbf B}
\def\bK{\mathbf K}
\def\Mv{\mathbf M}
\def\bM{\mathbf M}

\def\bD{\mathbf D}
\def\bS{\mathbf S}

\def\Sv{\mathbf S}

\def\bP{\mathbf P}

\def\bSigma{\mbox{\boldmath{$\Sigma$}}}

\def\balpha{\boldsymbol \alpha}
\def\bbeta{\boldsymbol \beta}
\def\tbalpha{\tilde\balpha}
\def\tbbeta{\tilde \bbeta}
\def\hbalpha{\hat \balpha}
\def\hbbeta{\hat \bbeta}
\def\trho{\tilde\rho}
\def\hrho{\hat\rho}

\def\smt{{\mbox{\tiny T}}}

\def\bone{{\bf 1}}
\def\bzero{{\bf 0}}

\newcommand{\Mc}{\mathcal{M}}
\def\Sc{\mathcal S}

\newcommand{\bgamma}{\mbox{\boldmath{${\bgamma}$}}}

\usepackage[figuresright]{rotating}

\author{Sandra E. Safo,  Jeongyoun Ahn, Yongho Jeon, and Sungkyu Jung \\
{ssafo@emory.edu, jyahn@uga.edu, yhjeon@yonsei.ac.kr, sungkyu@pitt.edu} \\
	Department of Biostatistics \& Bioinformatics, Emory University, Atlanta, GA\\
	Department of Statistics, University of Georgia, Athens, GA\\
	Department of Applied Statistics, Yonsei University, Seoul, South Korea\\
	Department of Statistics, University of Pittsburgh, Pittsburgh, PA}

\title{Sparse Generalized Eigenvalue Problem with Application to Canonical Correlation Analysis for Integrative Analysis of Methylation and Gene Expression Data}

\date{}

\begin{document}
	
	\maketitle
%	\label{firstpage}

\begin{abstract}
	We present a method for individual and integrative analysis of high dimension, low sample size data that capitalizes on the recurring theme in multivariate analysis of projecting higher dimensional data onto a few meaningful directions that are solutions to a generalized eigenvalue problem. We propose a general framework, called SELP (Sparse Estimation with Linear Programming), with which one can obtain a sparse estimate for a solution vector of a generalized eigenvalue problem. We demonstrate the utility of SELP on canonical correlation analysis for an integrative analysis of methylation and gene expression profiles from a breast cancer study, and we identify some genes known to be associated with breast carcinogenesis, which indicates that the proposed method is capable of generating biologically meaningful insights.
	Simulation studies suggest that the proposed method performs competitive in comparison with some existing methods in identifying true signals in various underlying covariance structures. 
\end{abstract}

%\textbf{Corresponding Author}: Sandra E. Safo; Email: ssafo@emory.edu

%\end{keywords}
\maketitle

\section{Introduction}\label{sec:intro}

Current advances in technology have led to the collection and processing of high dimension, low sample size (HDLSS) data, in which the number of measured variables is large relative to the number of experimental units. Increasingly, these data include multiple data types for the same experimental units. For instance, in biomedical research, these data types include gene expression, methylation, copy number variation, metabolomics data and many more. Since each of these measurements provides a different insight for the underlying  biological system, one could analyze the data individually, or more desirably, to analyze them simultaneously in order to study the association among different data types for a given subject \citep{Lock:2013,shen:2013}. We present a unified method for individual and integrative analysis of such data types to estimate an effective lower dimensional representation. The motivating example presented below is in a realm of the latter case, where canonical correlation analysis (CCA) is often chosen as an effective tool for integrative analysis of multiple data types.

\subsection{Holm Breast Cancer Study}\label{sec:data}

Breast cancer is the most commonly occurring cancer, and one of the leading causes of cancer deaths in women. Although it is well known that breast carcinogenesis is a multi-step process arising from genetic changes such as gene amplifications, mutations and copy number variations, recent studies also suggest that epigenetic factors such as DNA promoter methylation cause breast tumorigenesis by silencing tumor suppressor genes \citep{Dworkin:2009}.  DNA methylation is an epigenetic alteration that regulates gene expression and the maintenance of genomic structure, and an abnormal DNA methylation patterns have been associated with cancer and tumor suppressor genes.   Our work is motivated by the recent DNA methylation and gene expression analysis by \cite{Holm:2010} that identified methylation patterns in breast cancer molecular subtypes. 
In their study, raw methylation profiles from $189$ breast cancer samples were extracted using the Beadstudio Methylation Module (Illumina) for $1,452$ CpG sites  (corresponding to 803 cancer-related genes). $\beta$-values were stratified into three groups; $0$, $0.5$ and $1$ with $1$ values interpreted as hypermethylation. Relative methylation levels were then obtained by centering the stratified $\beta$-values across all samples. In addition, relative gene expression levels of $179$ out of $189$ breast cancer tumors were obtained using oligonucleotide arrays for $511$ probes. Thus $n=179$ will be the number of the common samples in the methylation and gene expression data. For the purpose of our analysis, we denote $\bX=[\bx_{1},\ldots,\bx_{p}]$, and $\bY=[\by_{1},\ldots,\by_{q}]$, $p=1,452$, and $q=511$, $\bx_{i}, \by_{j} \in \Re^{n}$, by the methylation and gene expression data matrices respectively.

The goal of our analysis is to jointly integrate methylation and gene expression data to investigate the overall dependency structure between CpG sites and genes using only a subset of the features.  This is a challenging task since the number of features $p$ or $q$ greatly exceeds the number of samples, $n$. Traditional multivariate approaches oftentimes yield results that lack interpretability, and are not able to identify important features. Our proposed method approaches this problem via a general framework for obtaining sparse estimation from a generalized eigenvalue problem, which will help us find a meaningful lower dimensional subspace of important CpG sites and genes in the high dimensional space that explain the overall dependency structure between the two data types.

\subsection{Generalized Eigenvalue Problem}\label{sec:gep}

The method we propose here is motivated by the recurring theme of many multivariate methods of projecting high dimensional data onto a meaningful, much lower dimensional subspace. A basis for this subspace can often be found by solving a generalized eigenvalue (GEV) problem.

The GEV problem for the pair of matrices $(\bM,\bS)$ is the problem of finding a pair $(\lambda, \bv)$ such that
\beq \label{eqn:gevp1}
\bM\bv = \lambda \bS\bv,
\eeq
where $\bM, \bS \in \Re^{p \times p}$ are usually symmetric matrices, $\bv \in \Re \setminus \{0\}$ and $\lambda \in \Re$. In most applications to statistical analyses, $\bS$ is positive or at least nonnegative definite and $\bM$ is often singular. The pair $(\lambda, \bv)$ that solves the GEV problem is called the generalized eigenvalue-eigenvector pair. Some popular and widely used data analysis methods that result from a GEV problem are principal component analysis (PCA), linear discriminant analysis (LDA), CCA, and multivariate analysis of variance (MANOVA). In PCA, principal components are directions with maximum variance of projected data. In LDA, the discriminant vectors are directions with  maximum separation between classes and a minimum variation within classes.
In CCA where the association between two sets of variables is of interest, the canonical correlation variables are determined by the directions of maximal correlation. Despite the popularity of these methods, one main drawback is the lack of sparsity. They have a limitation that their solution vector $\bv$ is a linear combination of all available variables, making it difficult to interpret the results oftentimes.

Sparse representations usually have physical interpretations in practice, and they have been shown to have good prediction performance in many high dimensional studies. Several approaches have been discussed to make $\hat\bv$ sparse.  A common approach is to apply sparse penalties such as lasso \citep{Tibshirani:1994}, adaptive lasso \citep{Zou:2006}, fused lasso \citep{TSRZK:2005}, elastic net \citep{ZH:2005} or SCAD \citep{FL:2001} to an objective function 
$\max_{\bv} \bv^{\smt}\bM\bv$, subject to the constraint $ \bv^{\smt}\bS \bv =1.$

In this paper, we propose a method to obtain a sparse estimate of $\bv$ in (\ref{eqn:gevp1}), called sparse estimation via linear programming (SELP). The primary benefits of our approach are two-folds. First, a general framework for obtaining a sparse solution to a GEV problem will be developed to allow for easier interpretation. We will also prove that the estimator is consistent when both $p$ and $n$ diverge.  We note that SELP may be applied to any GEV problem such as PCA, LDA, CCA, and  MANOVA. However, an actual application to a specific multivariate problem should be carefully carried out considering problem-specific challenges. For example, since every method has different goal,  a strategy for parameter tuning should be approached differently for each method. Second, we will implement SELP to CCA  for an integrative analysis of DNA methylation and gene expressions  to discover CpG sites and  genes that could shed light on the etiology of breast cancer. An efficient algorithm will be provided with a careful consideration on the tuning issue. 

%\subsection{Outline}
The rest of the paper is organized as follows. In Section \ref{sec:methods}, we present the SELP framework for obtaining sparse estimates from a generalized eigenvalue problem and prove a consistency of the proposed estimator.  In Section \ref{sec:subcca}, we develop a sparse CCA via SELP.  In Section \ref{sec:simul}, we conduct simulation studies to assess the performance  of our method under different settings and to compare with existing methods. In Section \ref{sec:real}, we apply our approach to the  motivating breast cancer study. We conclude with a discussion and remarks in Section \ref{sec:discuss}.

\section{Sparse Estimation by Linear Programming}\label{sec:methods}

Our proposed idea is motivated by the fact that one can obtain a sparse estimate of  a generalized eigenvector  by minimizing its $\ell_1$ norm while controlling maximum discrepancy in the equation (\ref{eqn:gevp1}). A similar idea has been proposed for the binary LDA by \cite{CL:2011}, which is inspired by the Dantzig selector (DS) \citep{Dantzig:2007}. \cite{Dantzig:2007} theoretically showed that the DS satisfies the oracle property of variable selection consistency and can be used as an effective variable selection method. Note that since the binary LDA is a regular eigenvalue problem, their direct estimation idea cannot be applied to a GEV estimation.

Motivated by the DS estimator, we consider the following optimization problem to the GEV problem (\ref{eqn:gevp1})
\bs
\min_{\bv} \|\bv\|_{1} \quad \mbox{subject to} \quad \|\bM\bv - \lambda\bS\bv \|_\infty \leq\tau, \; \lambda>0,
\es
where $\tau > 0$ is a tuning parameter. However, we call this a na\"ive approach,  since its solution is always the zero vector, which satisfies the $\ell_\infty$ constraint and has minimum $\ell_1$ norm. Thus we substitute one of $\bv$ in the constraint with a reasonable vector, which we choose to be the nonsparse eigenvector $\tilde{\bv}$ of $\bS^{-1}\bM$. Also we substitute the unknown $\lambda$ by the eigenvalue corresponding to $\tilde\bv$, denoted by $\tilde\lambda$. Then we propose to solve the following problem:
\beq\label{eq:submain2}
\min_{\bv} \|\bv\|_{1} \quad \mbox{subject to} \quad \|\bM\tilde{\bv} - {\tilde{\lambda}}\bS\bv \|_\infty \leq\tau_1,
\eeq
which we will call a Sparse Estimation via Linear Programming (SELP) approach for a generalized eigenvalue problem.
It is clear that when $\tau_1 =0$, the initial generalized eigenvector $\tilde{\bv}$ is recovered. Also note that one could alternatively use a constraint $ \|\bM{\bv} - {\tilde{\lambda}}\bS\tilde\bv \|_\infty \leq\tau_1$, however, we have found that this alternative often performs poorly due to the singularity of $\bM$. Furthermore, it is straightforward to see that this alternative does not necessarily recover the original nonsparse solution when $\tau_1 = 0$.

Once the solution to (\ref{eq:submain2}) is obtained, we can find sparse estimates of the subsequent generalized eigenvectors $\bv_2, \ldots, \bv_J$, where $J$ is the rank of $\bM$, by imposing an additional orthogonality condition.
For the $j$th vector, let ${\hat\bB}_{j}$ be a $p \times (j-1)$ matrix whose columns are $\hat{\bv}_{1},\ldots ,\hat{\bv}_{j-1}$.  Then for $\hat \bv_j$ we solve
\begin{equation}\label{eqn:multi}
\min_{\bv} \| \bv \|_{1} ~~\mbox{subject to}~~ \|\bM\tilde{\bv}_{j}-\tilde{\lambda}_{j}\bS\bv \|_{\infty} \leq \tau_{j} ~~\mbox{and}~~ {\hat\bB}_{j}^{\smt}\bS\bv  = 0, \nonumber
\end{equation}
where $(\tilde\lambda_j, \tilde\bv_j)$ is the $j$-th eigenvalue-eigenvector pair. Alternatively, one can project data onto the orthogonal complement of $\hat\bB_j$ then solve (\ref{eq:submain2}) without additional constraints. % How to tune the parameters $\tau_j$'s is an important issue in the implementation and should be considered according to the context of the problem.

\subsection{Consistency of SELP}\label{sec:theory}

Since the SELP method addresses a generalized eigenvalue problem, theoretical properties of the estimator mostly depend on the specific context that it is applied on. For example, if it is applied to multi-class LDA, then misclassification error would be one of the most important concern. In this section, however, we stay in the generality of the GEV problem and establish a consistency of our sparse estimator for a generalized eigenvector.

Let $\Mc$ be a $p \times p $ nonnegative definite matrix with rank($\Mc$) $ = m \le p $ and $\Sc$ be a $p \times p$ positive definite matrix. The true $\vv$ satisfies $\Mc\vv = \lambda \Sc \vv$. We assume $\vv$ is $s$-sparse for some fixed $s$, that is, the number of nonzero loadings of $\vv$ is $s$.
Let $\Mv$ and $\Sv$ be sample versions of $\Mc$ and $\Sc$ that preserve the same definiteness, respectively. Since $\Mv \tilde\vv = \tilde\lambda \Sv \tilde\vv$, we write the constraint of (\ref{eq:submain2}) as
\begin{equation} \label{eq:SELP-2}
\|  \Sv \tilde\vv - \Sv \vv \|_\infty \le  \tau_n
\end{equation}
for some $ \tau_n$. In what follows we show that the solution $\hat\vv$ to (\ref{eq:submain2}) is consistent with $\vv$ for both $p,n \to \infty$, if $ \tau_n = O( (\log p / n )^{1/2})$.

To this aim, let us assume that the sample versions are reasonable in the sense that with probability converging to 1 as $p$ and $n$ increases while $\log p / n \to 0 $,
\begin{equation} \label{eq:CriticalAssumption1}
\|\Sv - \Sc \|_{\rm{max}} \le c_1 \sqrt{\frac{\log p }{n}},
\end{equation}
and
\begin{equation} \label{eq:CriticalAssumption2}
\|\Sv\tilde\vv - \Sc \vv\|_{\infty} \le c_2 \sqrt{\frac{\log p }{n}},
\end{equation}
for some constants $c_1$ and $c_2$. Here $\|\bA\|_{max} = \max_{i,j} |a_{ij}|$ is the maximum absolute value of the elements. Whether the above two assumptions hold should be investigated for each statistical problem. For example in the binary classification setting, $\Sc$ and $\Sc \bv$ correspond to a common covariance matrix and mean difference vector respectively.  Under this setting, with common sample estimators,  it is known \cite{CL:2011} that both inequalities hold with a probability greater than $1- O(p^{-1})$ or $1- O(p^{-1} + n^{-\epsilon /8}) $ under mild distributional assumption such as sub-Gaussian tail or polynomial tail. Under these assumptions, and with some regularity conditions, we can show that the solution $\hat\vv$ is a consistent estimator of $\vv$, as described in the next theorem. Note that $\|\bA\|_1 = \max_j \sum_i |a_{ij}|$.

\begin{thm}
	Suppose the true vector $\vv$ is $s$-sparse, and that $\|\vv\|_2 = 1$. Assume that (\ref{eq:CriticalAssumption1}) and (\ref{eq:CriticalAssumption2}) hold with probability greater than $1 -  O(p^{-1})$, and that for some constant $M_0<\infty$, $\|\Sc^{-1}\|_1 \le M_0$. Then with probability at least $1 -  O(p^{-1})$, we have that
	\begin{align}
	\| \hat\vv - \vv \|_1 &\le  c s M_0 \tau_n \label{eq:thmL1} \\
	\| \hat\vv - \vv \|_2 &\le  c \sqrt{s} M_0 \tau_n \label{eq:thmL2} 
	\end{align}
	as long as $ \tau_n \ge c'\sqrt{\log p / n}$ in (\ref{eq:SELP-2}), and $c$, $c'$ depend only on $c_1,c_2$ and $s$.
\end{thm}

\section{Sparse Canonical Correlation Analysis} \label{sec:subcca}

\subsection{Canonical Correlation Analysis}
Suppose that we have two sets of random variables, $ \bsX = (X_{1},\dots,X_{p})^{\smt}$ and $\bsY = (Y_{1}\dots,Y_{q})^{\smt}$. Without loss of generality, we assume the variables have zero means. The goal of CCA \citep{Hotelling:1936} is to find linear combinations of the variables in $\bsX$, say $\bsX^{\smt} \balpha$ and in $\bsY$, say $\bsY^{\smt} \bbeta$ such that the correlation between these linear combinations is maximized. Let $\bSigma_{xx}$ and $\bSigma_{yy}$ be the population covariances of $\bsX$ and $\bsY$ respectively, and let $\bSigma_{xy}$ be the $p \times q$ matrix of  covariances  between $\bsX$ and $\bsY$. Let $\rho = \mbox{corr}(\bsX^{\smt} \balpha, \bsY^{\smt} \bbeta)$ be the correlation between the canonical correlation variables. Mathematically, the goal of CCA is to find $\balpha$ and $\bbeta$ that solves
\begin{equation}\label{eqn:cca}
\max_{\balpha, \bbeta} \mbox{corr}(\bsX^\smt\balpha, \bsY^\smt\bbeta) = \max_{\balpha,\bbeta}\frac{\balpha^{\smt}\bSigma_{xy}\bbeta}{\sqrt{\balpha^\smt \bSigma_{xx}\balpha} \sqrt{\bbeta^\smt \bSigma_{xx}\bbeta}}.
\end{equation}
The correlation coefficient in (\ref{eqn:cca}) is not affected by scaling of $\balpha$ and $\bbeta$, hence one can choose the denominator to be equal to one and solve the equivalent problem: Find $\balpha$ and $\bbeta$ that solves the optimization problem
\begin{eqnarray} \label{eqn:ccaopt}~~~~~~
\max_{\balpha,\bbeta} ~\balpha^{\smt}\bSigma_{xy}\bbeta ~~~\mbox{subject to}~~ \balpha^{\smt}\bSigma_{xx}\balpha =1~~ \mbox{and~~}\bbeta^{\smt}\bSigma_{yy}\bbeta =1.
\end{eqnarray}
Subsequent directions are obtained by imposing the following additional orthogonality constraints
\beq
\balpha^{\smt}_{i}\bSigma_{xx}\balpha_{j} = \bbeta^{\smt}_{i}\bSigma_{yy}\bbeta_{j} =\balpha^{\smt}_{i}\bSigma_{xy}\bbeta_{j}=0 ,~~ i \neq j, ~~i,j=1,\dots,\min (p,q).\nonumber
\eeq
Using Lagrangian multipliers $\rho$ and $\mu$ on (\ref{eqn:ccaopt}),  we have
\begin{equation}\label{eqn:ccalag}
L(\balpha,\bbeta, \rho,\mu) = \balpha^{\smt}\bSigma_{xy}\bbeta- (\rho/2)(\balpha^{\smt}\bSigma_{xx}\balpha -1) - (\mu/2)(\bbeta^{\smt}\bSigma_{yy}\bbeta -1).
\end{equation}
Differentiating (\ref{eqn:ccalag}) with respect to $\balpha$ and $\bbeta$ and setting the derivatives to zero yields
\begin{eqnarray}
\frac{\partial L}{\partial \balpha} &=&\bSigma_{xy}\bbeta - \rho ~\bSigma_{xx}\balpha = 0;\label{ccalag1}\\
\frac{\partial L}{\partial \bbeta} &=&\bSigma_{yx}\balpha - \mu~\bSigma_{yy}\bbeta= 0\label{ccalag2}.
\end{eqnarray}
Note that pre-multiplying equations (\ref{ccalag1}) and (\ref{ccalag2}) by $\balpha^{\smt}$ and $\bbeta^{\smt}$ respectively results in $\rho=\mu$. Equations (\ref{ccalag1}) and (\ref{ccalag2}) may be jointly re-written in the form of the GEV problem of (\ref{eqn:gevp1})
\begin{equation} \label{eq:ccagep1}
\left[\begin{array}{cc}
0 & \bSigma_{xy} \\
\bSigma_{yx} & 0
\end{array}  \right]  \left[\begin{array}{c}
\balpha  \\
\bbeta
\end{array}  \right] = \rho \left[\begin{array}{cc}
\bSigma_{xx} & 0 \\
0 & \bSigma_{yy}
\end{array}  \right]\left[\begin{array}{c}
\balpha  \\
\bbeta
\end{array}  \right],
\end{equation}
which can be solved by applying singular value decomposition (SVD) \citep{MKB:2003} to the matrix
\begin{equation}\label{eqn:svdK}
\bK = \bSigma_{xx}^{-1/2}\bSigma_{xy}\bSigma_{yy}^{-1/2} = (\bu_{1},\ldots,\bu_{k})\bD(\bv_{1},\ldots,\bv_{k})^{\smt}.
\end{equation}
Here, $k$ is the rank of the matrix $\bK$, $\bu_{j}$ and $\bv_{j}$, ($j=1,\ldots,k$) are the $j$th left and right singular vectors of $\bK$, and $\bD$ is a diagonal matrix containing singular values $\lambda_{j}$ of $\bK$ ordered from the largest to the smallest. It follows that the $j$th canonical variables can be obtained by
\begin{eqnarray} \label{nonsparse}
\tilde{\balpha}_{j} = \bSigma^{-1/2}_{xx}\bu_{j}, ~\tilde{\bbeta}_{j} = \bSigma^{-1/2}_{yy}\bv_{j},
\end{eqnarray}
and the $j$th canonical correlation is $\tilde\rho_j = \lambda_j$.
In practice, $\bSigma_{xx}^{-1/2}\bSigma_{xy}\bSigma_{yy}^{-1/2}$ is replaced with the usual sample version $\bS_{xx}^{-1/2}\bS_{xy}\bS_{yy}^{-1/2}$, which results in consistent estimators of $\balpha$ and $\bbeta$ for fixed dimensions $p, q$ and large sample size $n$.  When  $p$ or $q$ are greater than $n$, a certain type of regularization is desired to avoid overfitting and singularity problem.

%\subsection{Some Sparse CCA Methods}

To enhance interpretability for high dimensional applications, various sparse CCA methods have been proposed. Most of these works achieve sparsity by adding penalty functions of the canonical correlation vectors in  (\ref{eqn:ccaopt})  \citep{WTH:2009} or its variant \citep{Maetal:2015}, or by thresholding $\bK$ in (\ref{eqn:svdK}) after right and left multiplication by previous estimates of canonical vectors \citep{PTB:2009, FL:2001, Chenetal:2013}.  \citet{CF:2012} compared several sparsity penalty functions such as lasso \citep{Tibshirani:1994}, elastic net \citep{ZH:2005}, SCAD \citep{FL:2001} and hard-thresholding, based on the algorithm of \citet{PTB:2009}. They conclude that elastic net and SCAD tend to yield a higher canonical correlation while maintaining sparsity.

\subsection{SELP for Sparse CCA} \label{sec:proposed}

In this section, we develop sparse CCA  via the GEV optimization (\ref{eq:submain2}). First we replace the population covariance matrices in (\ref{eq:ccagep1}) by sample estimates that are reasonable for high dimensional data. 
Some existing works regularize the covariance matrices by assuming that within covariances are diagonal, which means identity for standardized data \citep{PTB:2009, WTH:2009,FL:2001}. Recently, \citep{Chenetal:2013} imposed assumptions such as sparsity, bandable, and Toeplitz on the covariance matrices.  However, \citep{Maetal:2015} reported that such structural assumptions may not be necessary and did not impose any restrictive assumptions on within covariances. In this work we consider two approaches. The first is to apply a small number to the diagonals: 
\beq\label{eq:ridge}
\tilde{\bS}_{xx}=\bS_{xx} + \sqrt{\log{p}/n}\bI, ~~\tilde{\bS}_{yy} =\bS_{yy} + \sqrt{\log{q}/n}\bI.
\eeq
Here the ridge coefficient $\sqrt{\log p /n}$ is a common choice for high dimensional regularization studies \cite{CL:2011}, and often assumed to diminish. We also consider a diagonal covariance matrix, as often assumed by many existing approaches, i.e., $\tilde\bS_{xx} = \bI_p$, and $\tilde\bS_{yy} = \bI_q$.

Now the CCA can be written as a sample version of the GEV problem of (\ref{eq:ccagep1}):
\beq \label{eq:ccaselp}
\left[\begin{array}{cc}
	0 & \bS_{xy} \\
	\bS_{yx} & 0
\end{array}  \right]  \left[\begin{array}{c}
\balpha  \\
\bbeta
\end{array}  \right] = \rho \left[\begin{array}{cc}
\tilde\bS_{xx} & 0 \\
0 & \tilde \bS_{yy}
\end{array}  \right]\left[\begin{array}{c}
\balpha  \\
\bbeta
\end{array}  \right],
\eeq
and the following formulation can be considered to obtain a sparse solution to the GEV problem (\ref{eq:ccaselp}) using ideas in Section \ref{sec:methods}:
\beq\label{eq:submain}
\min_{\balpha,\bbeta} 
\left\| \! \left[ \! \begin{array}{c}
	\!\balpha \! \\
	\!\bbeta\!
\end{array} \! \right] \!\right\|_{1} 
~\mbox{subject to}~ \left\| \!   \!
\left[ \! \begin{array}{cc}
	\!0            \!&\! \bS_{xy} \!\\
	\!\bS_{yx} \!&\! 0\!
\end{array}  \! \right]
\!
\left[ \!\begin{array}{c}
	\tilde{\!\balpha}  \!\\
	\tilde{\!\bbeta}\!
\end{array} \!\right]\!
- {\tilde{\rho}}
\left[\!\begin{array}{cc}
	\!\tilde\bS_{xx} \!&\! 0 \!\\
	\!0 \!&\! \tilde \bS_{yy}\!
\end{array} \! \right]\!
\!
\left[ \!\begin{array}{c}
	\!\balpha  \!\\
	\!\bbeta\!
\end{array} \!\right]
\right\|_{\infty} \! \leq\tau,
\eeq
where $\tilde{\balpha}$ and $\tilde{\bbeta}$ are the (nonsparse) solution to (\ref{eq:ccaselp}) and $\tilde\rho$ is the corresponding eigenvalue obtained from (\ref{nonsparse}). Let ($\tilde\balpha_1$, $\tilde\bbeta_1$) be the first  (nonsparse) solution to (\ref{eq:ccaselp}) and $\tilde\rho_{1}$ the corresponding eigenvalue, which can be calculated in $O(n^2p)$ instead of $O(p^3)$ using the transformation discussed in \citep{HT:2004}. Applying SELP from (\ref{eq:submain2}), we solve the problem (\ref{eq:submain}) using the following two optimization problems:
\begin{eqnarray}\label{eq:ccafirst1}
\min_{\balpha} \|\balpha\|_{1} \quad \mbox{subject to} \quad \|\bS_{xy}\tilde{\bbeta_1}- {\tilde{\rho}_{1}}\tilde{\bS}_{xx}\balpha \|_\infty \leq\tau_{x},\\
\label{eq:ccafirst2}
\min_{\bbeta} \|\bbeta\|_{1} \quad \mbox{subject to} \quad \|\bS_{yx}\tilde{\balpha_1}- {\tilde{\rho}_{1}}\tilde{\bS}_{yy}\bbeta \|_\infty \leq\tau_{y}.
\end{eqnarray}
for $\tau_{x}, \tau_{y} \ge 0$. We can find $\hat{\balpha}_{j}$ and $\hat{\bbeta}_{j}$, $j \ge 2$, by solving (\ref{eq:ccafirst1}) and (\ref{eq:ccafirst2}) after projecting data onto the orthogonal complement of $[\hat\balpha_1, \ldots, \hat\balpha_{j-1}]$ and $[\hat\bbeta_1, \ldots, \hat\bbeta_{j-1}]$, respectively. In other words, we `deflate' data by obtaining $\bX_{new} = \bX \bP_j^\perp$, where $\bP_j^\perp$ is the projection matrix onto the orthogonal complement of  $[\hat\balpha_1, \ldots, \hat\balpha_{j-1}]$. Similar for $\bY$. 

\subsection{Implementation}\label{subsec:algorithm} We first normalize the columns of $\bX$ and $\bY$ to have mean zero and unit variance. The objective and constraint functions  of the problems (\ref{eq:ccafirst1}) and (\ref{eq:ccafirst2})  are linear, allowing us to solve  the problem via linear programming.
For  (\ref{eq:ccafirst1}),  let $\alpha_{j} = \alpha_{j}^{+} - \alpha_{j}^{-}$, $j=1,\ldots,p$, with $\alpha_{j}^{+}, \alpha_{j}^{-} \ge 0$, $\balpha^{+}=(\alpha_1^{+},\dots,\alpha_p^{+})^\smt$, $\balpha^{-}=(\alpha_1^{-},\dots,\alpha_p^{-})^\smt$, $\tilde{\bS} = {\tilde{\rho}_{1}}\tilde{\bS}_{xx}$, and $\bell=\bS_{xy}\tilde{\bbeta_1}$. Then, $|\alpha_{j}|= \alpha_{j}^{+} + \alpha_{j}^{-}$ and the problem (\ref{eq:ccafirst1}) is written as
\begin{eqnarray*}
	\min_{\balpha^{+},\balpha^{-}} ~~ \displaystyle\sum\limits_{j=1}^p \alpha_{j}^{+} + \sum\limits_{j=1}^p \alpha_{j}^{-} ~~~~~  \mbox{subject to}\nonumber
	~~~\begin{array}{c}
		\balpha^{+} \geq \bzero_p, ~~~\balpha^{-}\geq \bzero_p, \\
		\tilde{\bS}(\balpha^{+} - \balpha^{-}) \leq \bone_{p}\tau_x + \bell,   \\
		-\tilde{\bS}(\balpha^{+} - \balpha^{-}) \leq \bone_{p}\tau_x - \bell,  \\
	\end{array}
\end{eqnarray*}
where the inequalities are element-wise. This can be solved by any off-the-shelf linear programming software.  The problem (\ref{eq:ccafirst2}) is solved in a similar way.

Since the proposed method uses the nonsparse solution ($\tilde\balpha_j, \tilde\bbeta_j, \tilde\rho_j)$ as the `initial' values, it is possible that the  effectiveness of the proposed method can be dependent on the quality of initial values. To alleviate the dependence we propose to iterate the procedure by updating the $(\tilde\balpha_j, \tilde\bbeta_j, \tilde\rho_j)$ with the found $(\hat\balpha_j, \hat\bbeta_j, \hat\rho_j)$ until convergence. Here $\hat\rho_j$ is the correlation coefficient between $\bX\hat\balpha_j$ and $\bY\hat\bbeta_j$. In all our empirical studies, the procedure reached convergence (the $\ell_2$ difference between successive solutions $<10^{-5}$) within $4\sim 5$ iterations. Algorithm 1 below describes the procedure to obtain $\hat\balpha_j$ and $\hat\bbeta_j$, $j = 1, \ldots, J$.

\begin{algorithm}[htbp!]
	\caption{Sparse CCA vectors via SELP}
	\begin{algorithmic}[1]
		\For{$j = 1, \ldots, J$}
		\State Standardize all variables. Initialize  with nonsparse estimates: $\tilde{\balpha}_{j}=\tilde{\bS}^{-1/2}_{xx}\bu_{1}$, $\tilde{\bbeta}_{j}=\tilde{\bS}^{-1/2}_{yy}\bv_{1}$, where $\bu_1$ and $\bv_1$ are the first left and right singular vectors in (\ref{eqn:svdK}). Also $\tilde\rho_j = \lambda_{1}^{1/2}$. Normalize $\tilde\balpha_j$ and $\tilde\bbeta_j$.
		\For{until convergence or some maximum number of iterations}
		
		\State Find $\hat{\balpha}_{j}$ and $\hat{\bbeta}_{j}$, by solving
		\[
		\min_{\balpha} \|\balpha\|_{1} \quad \mbox{subject to}
		\quad \|{\bS}_{xy}\tilde{\bbeta}_{j}- {\tilde{\rho}_{j}}\tilde{\bS}_{xx}\balpha \|_\infty \leq\tau_{x}
		\]
		\[
		\min_{\bbeta} \|\bbeta\|_{1} \quad \mbox{subject to}
		\quad \|\bS_{yx}\tilde{\balpha}_{j}- {\tilde{\rho}_{j}}\bS_{yy}\bbeta \|_\infty \leq\tau_{y}
		\]
		\State Normalize $\hat{\balpha}_{j}$ and $\hat{\bbeta}_{j}$ to have unity $l_{2}$ norm and obtain the canonical correlation coefficient $\hat{\rho}_{j}$.
		\State Update ($\tbalpha_j, \tbbeta_j, \trho_j$)  with ($\hbalpha_j, \hbbeta_j, \hrho_j$).
		\EndFor
		\State If $j \ge 2$, update $\bX$ and $\bY$ by projecting them to the orthogonal complement of $[\hbalpha_1, \ldots, \hbalpha_{j-1}]$ and $[\hbbeta_1, \ldots, \hbbeta_{j-1}]$ respectively.
		
		\EndFor
	\end{algorithmic}
\end{algorithm}

\subsection{Selection of tuning parameters}\label{subsec:tuning}

The tuning parameters $\tau_x$ and $\tau_y$ control the degree of sparsity of the solution vectors. A near zero value will yield a nonsparse solution while $\tau_x =  \|\bS_{xy}\tbbeta_j\|_{\infty}$ or $\tau_y = \|\bS_{yx}\tbalpha_j\|_\infty$ will yield null vectors, which will give us a natural upper bound for tuning. We suggest to choose the tuning parameters from $(0, \|\bS_{xy}\tbbeta_j\|_{\infty})$ for $\tau_x$ and $(0,  \|\bS_{yx}\tbalpha_j\|_\infty)$ for $\tau_y$ via a $V$-fold cross-validation.

Determining a cross-validation criterion in CCA is not straightforward, since there is no notion of `prediction'. One might naively choose the value of achieved canonical correlation $\hrho$. However, this will almost always prefer a smaller $\tau$, which yields a less sparse vector.
In this work we propose to use the following criterion, which measures the stability of the solution. We randomly group the rows of $\bX$ and $\bY$ into $V$ roughly equal-sized groups, denoted by $\bX^{1},\ldots,\bX^{V}$, and $\bY^{1},\ldots,\bY^{V}$, respectively. For $v=1,\ldots,V$, let $\bX^{-v}$ and $\bY^{-v}$ be the data matrix leaving out $\bX^{v}$ and $\bY^{v}$, respectively.
For given $(\tau_{x}, \tau_{y})$, we apply Algorithm 1 on $\bX^{-v}$ and $\bY^{-v}$ to derive the canonical correlation vectors $\hat{\balpha}^{-v}_{j}(\tau_{x}, \tau_{y})$,  and $\hat{\bbeta}^{-v}_{j}(\tau_{x}, \tau_{y}),  j=1,\ldots, J$. The sample  canonical correlation coefficients for the training and testing data are obtained as
$\hat{\rho}_{j_{train}}^{-v} = \text{corr}( \bX^{-v}\hat{\balpha}^{-v}_{j}, \bY^{-v}\hat{\bbeta}^{-v}_{j})$ and
$\hat{\rho}_{j_{test}}^{v}= \text{corr}( \bX^{v}\hat{\balpha}^{-v}_{j}, \bY^{v}\hat{\bbeta}^{-v}_{j})$, and the optimal $(\tau_{x}, \tau_{y})$ are selected so that they  minimize
\begin{equation}\label{eqn:tunsel2}
CV(\tau_{x}, \tau_{y})=\bigg (\sum\limits_{v=1}^{V}\left|\hat{\rho}_{j_{train}}^{-v} (\tau_{x}, \tau_{y})\right|- \sum\limits_{v=1}^{V}\left|\hat{\rho}_{j_{test}}^{v} (\tau_{x}, \tau_{y})\right|\bigg)^2.
\end{equation}

We note that the tuning parameter criterion (\ref{eqn:tunsel2}) is motivated by the approach of \citep{PTB:2009} which minimizes the average difference between the canonical correlation of the training and testing sets. However, a potential drawback of their approach is that there may be a lot of variability in the $V$ correlation estimates since the correlations from the training set are mostly higher than the correlations from the testing sets. Therefore, we adopt a more natural measure that leverages the variability in the average correlation by minimizing over the differences between the average canonical correlations from the training and testing sets.

The optimal tuning parameter pair can be  chosen by performing a grid search over a pre-specified set of parameter values. In our empirical studies, in order to  reduce computational costs, we fix $\tau_{y}$ at some value, say  the midpoint of the grid and search for the optimal value for $\tau_x$. Then we fix $\tau_x$ at that value and search for the best $\tau_y$. Lastly, we apply the chosen tuning parameters to the whole data to find the final estimates.  The process is repeated for each $j$.

\section{ Simulation studies}\label{sec:simul}
We conduct Monte Carlo simulations to assess the performance of the proposed methods in comparison with some existing methods.
We generate $(p + q)-$ dimensional random variable $\bZ = (\bX^\smt, \bY^\smt)^\smt$ from multivariate normal with zero mean and covariance $\bSigma$, which is partitioned as
\bs
\bSigma =\left(
\begin{array}{cc}
	\bSigma_{xx} & \bSigma_{xy} \\
	\bSigma_{yx} & \bSigma_{yy}
\end{array} \right),
\es
where $\bSigma_{xy}$ is the covariance between $\bX$ and $\bY$, and $\bSigma_{xx}$, $\bSigma_{yy}$ are respectively the covariance of $\bX$ and $\bY$.

With the common sample size $n = 80$, the dimension $p = 200$ in all three settings and $q = 150$ in Settings 1 and 2, while $q = 200$ in Setting 3.
We fix the number of signal variables in either set to be $1/10$ of the whole set, i.e., there are 20 signal variables in the $\bX$-set where there are 15 or 20 in the $\bY$-set. In all three settings the within-covariance is block-diagonal, so that  noise variables are not correlated with signal variables within respective sets. In both Settings 1 and 2, the signal variables have within-set correlation .7 and between-set correlation .6. The difference between the two settings is that noises are uncorrelated in Setting 1, but mildly correlated ($.1$) in Setting 2. This may be the case in many biomedical studies where genomic data are correlated within a pathway and uncorrelated between pathways. Note that the first two settings have only one true canonical pair in the population structure, thus estimating the first pair ($\balpha_1, \bbeta_1$) is sufficient.  In Setting 3, we consider a case when there are two underlying canonical pairs. Similar to a setting in \citet{Chenetal:2013}, we set the between covariance $\Sigma_{xy}$ as $\bSigma_{xx} \bA\bD\bB^\smt\bSigma_{yy}$, where $\bD = \text{diag}(.9, .6)$ has two population canonical correlation coefficients, and $\bA =[\balpha_1, \balpha_2]$ and $\bB = [\bbeta_1, \bbeta_2]$ are matrices of corresponding canonical vectors. The two canonical pairs are:
\[
\balpha_{1}=\bbeta_{1} \propto   \left(-\mathbf{1}_{10}, \mathbf{0}_{p-10} \right)^{\smt}, \hspace{.3in}
\balpha_{2}=\bbeta_{2}\propto \left( \mathbf{0}_{10}, \mathbf{1}_{10}, \mathbf{0}_{p-20}\right)^{\smt}.
\]
Table \ref{tab:simulset} summarizes the simulation settings. We generate $100$ Monte Carlo datasets, within which we use 5-fold cross validation to select the tuning parameters and then use the whole dataset to obtain canonical estimates. 
\begin{table}[!htbp]
	\caption{Description of simulation settings. {BD} denotes Block Diagonal and {CS}$(\rho)_m$ denotes $m-$dimensional compound symmetry with diagonal one and off-diagonal $\rho$. $\bJ$ is a matrix of ones.\label{tab:simulset}}
	\begin{tabular}{c||c|c|c}
		\hline
		Setting & $\bSigma_{xx}$ & $\bSigma_{yy}$ & $\bSigma_{xy}$ \\
		\hline 
		1 &  BD[CS$(.7)_{20}, \bI_{180}$] & BD[CS$(.7)_{15}, \bI_{135}$] &  BD[$\bJ_{20\times 15}, \bzero_{180 \times 135}]$ \\
		\hline
		2 &  BD[CS$(.7)_{20}$, CS$(.1)_{180}$] & BD[CS$(.7)_{15}$, CS$(.1)_{135}$] &  BD[$\bJ_{20\times 15}, \bzero_{180 \times 135}$] \\						
		\hline
		3 &   BD[CS$(.7)_{20}, \bI_{180}$]& BD[CS$(.7)_{20}, \bI_{180}$] &  $\bSigma_{xx}\bA\bD\bB^\smt\bSigma_{yy}$ \\ 
		\hline
	\end{tabular}
\end{table}

We compare the proposed method with the following existing methods: sparse CCA via Covex Optimization with group-Lasso Refinement (CoLaR) \citep{Maetal:2015}, sparse CCA (SCCA) \citep{PTB:2009}, penalized matrix decomposition CCA with lasso penalties (PMD) \citep{WTH:2009} and sparse CCA with SCAD penalty (SCAD) \citep{CF:2012}.  Since all  these approaches except CoLaR assume identity within-set covariance, we  implement the proposed SELP approach in two different ways. We will call the proposed method that uses identity for within-covariance as SELP-I, while SELP-R  is for the method using the ridge-corrected within-covariance in (\ref{eq:ridge}). Note that we did not compare our methods with  \citet{Chenetal:2013} and \citet{Chaoetal2:2015} because there were no software available to implement these methods. We implement CoLaR using the MATLAB package SCCALab \citep{Maetal:2015}, SCCA using the R code from the authors website \citep{SCCA2:2009}, PMD using the R-package PMA \citep{PMA:2013}, and SCAD using the R code \citep{PTB:2009} with SCAD penalty. We implement SELP using the convex optimization software CVX \citep{cvx:2012}. 
The tuning parameters for each method are chosen based on their own tuning strategies. Briefly, CoLaR, SCCA and CCA with SCAD choose the optimal tuning parameter pair that maximizes the average correlation vectors in the testing set. PMD chooses the optimal tuning parameter pair using a permutation scheme. See \cite{PMA:2013} for details. 

We evaluate these methods using the following criteria.
\begin{enumerate}
	\item \textit{Estimation accuracy}: We measure the error in estimating the subspace spanned by canonical vectors, i.e., 
	$\|\hbalpha\hbalpha^{\smt} - \balpha\balpha^\smt\|^2_{F}$ and  $\|\hbbeta\hbbeta^{\smt} -\bbeta\bbeta^{\smt}\|^2_{F}$, where $\|\cdot\|_F$ is the Frobenius norm. 
	
	\item \textit{Selectivity}: We consider three measures for selectivity of relevant features: sensitivity, specificity, and Matthew's correlation coefficient defined as follows: 
	\begin{eqnarray}
	\mbox{Sensitivity}&=&\frac{TP}{TP + FN}, \nonumber \\
	\mbox{Specificity}&=&\frac{TN}{TN + FP}, \nonumber \\
	\mbox{MCC} &=& \frac{TP \cdot TN - FP \cdot FN}{\sqrt {(TP + FN)(TN + FP)(TP + FP)(TN + FN)}},\nonumber
	\end{eqnarray}
	where TP, FP, TN, FN are true positives, false positives, true negatives, and false negatives respectively. We  note that  MCC lies in the interval $[-1,1]$,  where a value of $1$ correspond to selection of all signal variables and no noise variables, a perfect selection. A value of $-1$ indicates total disagreement between signal and noise, and a value of $0$ implies random guessing.\\
	
	\item \textit{Canonical correlation coefficient}: The third comparison criterion is the estimated canonical correlation that maximizes the association between $\bX$ and $\bY$ and  is given by
	$\hat{\rho} = \text{Corr} (\bX\hat{\balpha}, \bY\hat{\bbeta})$.
\end{enumerate}

\subsection{Simulation Results}
\begin{figure}[htbp!]
	\centering
	\includegraphics[height = 2.2in]{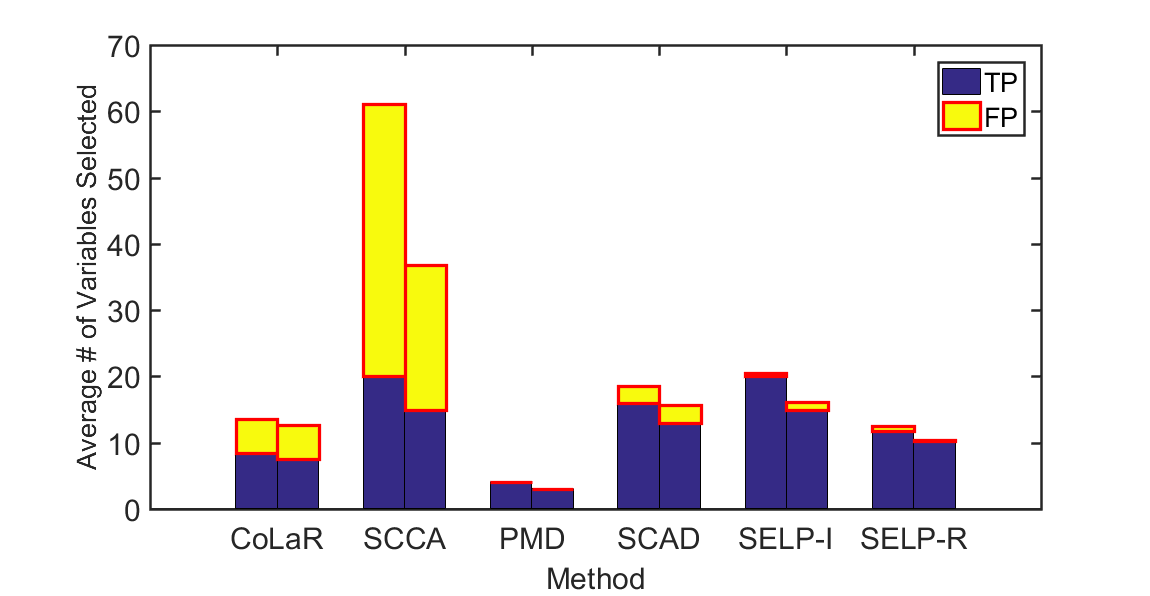} \\
	(a) Setting 1 \\ 
	\includegraphics[height = 2.2in]{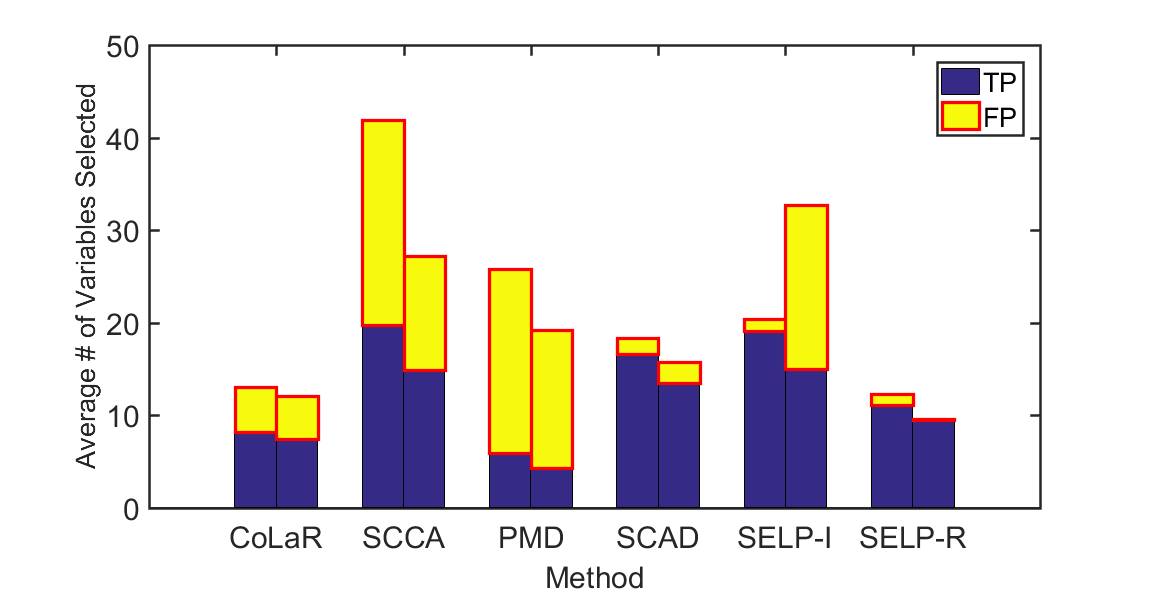}  \\ 		
	(b) Setting 2 \\ 
	\includegraphics[height = 2.3in]{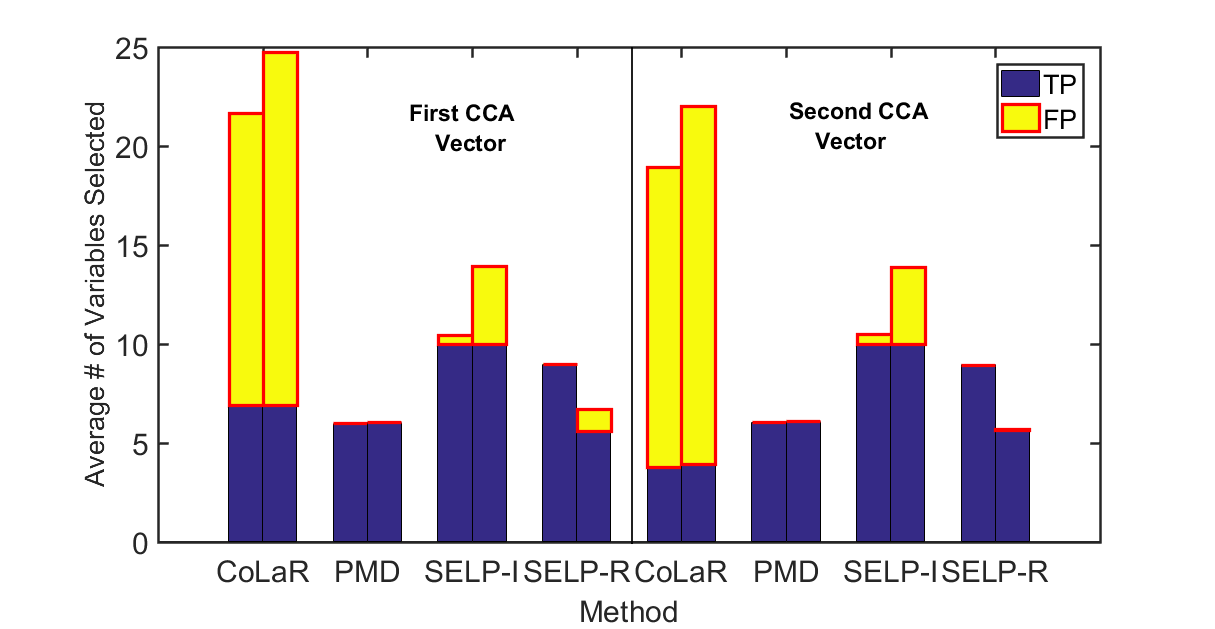}\\
	(c) Setting 3 
	\caption{Selectivity results from simulation study. Each pair of bars shows the number of selected variables for $(\hbalpha, \hbbeta)$. Each bar is divided into two colors, where the upper portion indicates falsely selected variables and lower portion is for true signal variables selected.  True number of signals is $(\balpha,\bbeta)=(20,15)$ in Settings 1 and 2, and $(10, 10)$ in Setting 3 for both first and second pairs. \label{fig:simresults}}
\end{figure}

Tables \ref{tab:simulresult1} - \ref{tab:simulresult3} show the average of the evaluation measures from 100 repetitions, from Settings 1 - 3 respectively. Note that the true canonical correlation coefficients for three settings are $0.84$, $0.84$ and $(0.9, 0.6)$ respectively. We observe that most methods estimate $\rho$ reasonably well for all three settings. All three tables suggest that the proposed method, especially SELP-I, performs competitively. The superior performance of the proposed method is also highlighted by the variable selection plots in Figure \ref{fig:simresults}. These plots show the number of variables selected by the methods, which is the height of each bar, divided into the portion of signal (TP) and noise (FP) variables. We observe that SELP-I selects the correct number of signals in general, but has more FP than SELP-R, yet lower than some sparse methods. On the other hand, SELP-R tends to be more sparse. For other sparse methods, SCCA and SCAD tend to identify all true signals but the former has more FP, while PMD selects less signals in all settings but has large FP in Setting 2. In Setting 3 where there are two true CCA vectors,  we show results for SELP, CoLaR and PMD in Table \ref{tab:simulresult3} since SCCA and SCAD do not produce multiple CCA vectors. We observe that SELP has higher sensitivity and MCC than CoLaR and PMD, comparable specificity estimates with PMD, and does better in estimating the true CCA vectors. The performance of CoLaR is suboptimal in terms of specificity, estimation error, and MCC.  Simulation results imply that assuming diagonal covariances within each dataset for SELP may result in better performance. 
\begin{table}[!htbp]
	\caption{Simulation results from Setting 1.  Best results in each row are highlighted.\label{tab:simulresult1}}
	\centering
	\begin{tabular}[htdp!]{cc|cccccc}
		\hline
		~&~&CoLaR	&SCCA&	PMD	&SCAD	&SELP-I&	SELP-R\\						
		\hline				
		\multirow{4}{*}{$\hat\balpha_1$}&Estimation error &1.227		&0.262	&1.323	&0.471	&\textbf{0.154}	&1.120\\
		~&Sensitivity&0.423			&\textbf{1.000}	&0.206	&0.802	&\textbf{1.000}	&0.587\\
		&Specificity&0.971			&0.772	&\textbf{1.000}	&0.986	&0.997	&0.996\\
		~&MCC&0.470			&0.503	&0.435	&0.816	&\textbf{0.987}	&0.721\\
		\hline					
		%~&~& ~&~ &~ &~ &~\\				
		\multirow{4}{*}{$\hat\bbeta_1$}&Estimation error&1.187		&0.222	&1.323	&0.394	&\textbf{0.153}	&1.033\\
		~&Sensitivity&0.501		&\textbf{1.000}	&0.203	&0.863	&\textbf{1.000}	&0.683\\
		&Specificity&0.962			&0.839	&\textbf{1.000}	&0.980	&0.991	&0.999\\
		~&MCC&0.499			&0.585	&0.432	&0.828	&\textbf{0.959}	&0.804\\ \hline
		~&$\hat{\rho}$&0.908			&0.858	&0.814	&0.838	&\textbf{0.839}	&0.877\\
		\hline
	\end{tabular}
\end{table}

\begin{table}[!htbp]
	\caption{Simulation results from Setting 2. Best results in each row are highlighted. \label{tab:simulresult2}	}
	\centering
	\begin{tabular}{cc|cccccc}
		\hline
		~&~&CoLaR	&SCCA&	PMD	&SCAD	&SELP-I&	SELP-R\\
		\hline				
		\multirow{4}{*}{$\hat\balpha_1$}&Estimation error &1.223			&\textbf{0.222}	&1.240	&0.456	&0.288	&1.146\\
		&Sensitivity&0.413				&\textbf{0.991}	&0.298	&0.833	&0.956	&0.559\\
		&Specificity&0.974	&0.877	&0.890	&0.991	&0.993	&\textbf{0.994}\\
		&MCC&0.470				&0.639	&0.168	&0.855	&\textbf{0.939}	&0.689\\
		%&~&~& ~&~ &~ &~ \\					
		\hline
		\multirow{4}{*}{$\hat\bbeta_1$}&Estimation error &1.185			&\textbf{0.200}	&1.240	&0.364	&0.236	&1.067\\
		& Sensitivity&0.499				&0.991	&0.290	&0.898	&\textbf{0.999}	&0.632\\
		&Specificity&0.966	&0.908	&0.890	&0.983	&0.868	&\textbf{0.999}\\
		&MCC&0.513				&0.700	&0.162	&0.862	&0.630	&\textbf{0.773}\\
		\hline
		&$\hat{\rho}$&0.906				&0.850	&0.824	&\textbf{0.842}	&0.849	&0.873\\
		\hline
	\end{tabular}
\end{table}

\begin{table}[!htbp]
	\caption{Simulation results from Setting 3. Best results in each row are highlighted.\label{tab:simulresult3}}	
	\centering
	\begin{tabular}[htdp!]{cc|cccc|cccc}
		\hline
		& & \multicolumn{4}{c|}{First pair} & \multicolumn{4}{c}{Second pair}  \\
		&~&CoLaR &PMD&	SELP-I	&SELP-R&CoLaR  &	PMD	&SELP-I 	&SELP-R	\\
		\hline
		\multirow{4}{*}{$\hat\balpha$}&Estimation error &1.401		&1.072	&\textbf{0.114}	&0.778&1.4061	&1.072	&\textbf{0.112}	&0.774\\	
		&Sensitivity&0.6930			&0.600	&\textbf{1.000}	&0.898&0.380	&0.600	&\textbf{1.000}	&0.895\\	
		&Specificity&0.920 &\textbf{1.000}	&0.998	&\textbf{1.000}&0.920	&\textbf{1.000}	&0.997	&\textbf{1.000}\\	
		&MCC&0.432			&0.767	&\textbf{0.976}	&0.943&0.201	&0.771	&\textbf{0.975}	&0.692\\	
		\hline
		\multirow{4}{*}{$\hat\bbeta$}&Estimation error &1.400		&1.072	&\textbf{0.324}	&1.118&1.402	&1.075	&\textbf{0.314}	&1.113\\	
		&Sensitivity&0.6900			&0.605	&\textbf{1.000}	&0.559&0.393	&0.605	&\textbf{1.000}	&0.565\\	
		&Specificity&0.906 &\textbf{1.000}	&0.979	&0.994&0.905	&\textbf{1.000}	&0.980	&0.996\\	
		&MCC&0.425			&0.769	&\textbf{0.839}	&0.670&0.207	&0.770	&\textbf{0.839}	&0.699\\	
		\hline
		&$\hat{\rho}$&0.935	        & 0.890	&\textbf{0.903}	&0.918&0.793	&0.657	&\textbf{0.646}	&0.692\\	
		\hline	\end{tabular}
\end{table}

\section{Analysis of the Motivating Data} \label{sec:real}
\subsection{Implementation}
We apply the proposed method to jointly analyze the methylation and gene expression data from the motivating Holm breast cancer study. There are $1,452$ CpG sites (corresponding to $803$ unique cancer-related genes), and $511$ probes on $179$ samples. Following the preprocessing step taken in the original Holm study, we filtered the methylation data so as to include the most variable methylated sites by removing CpG sites with standard deviation less than 0.3 across samples. This resulted in $334$ CpG sites corresponding to $249$ cancer-related genes. The gene expression data were used without preprocessing as done in the Holm study. We identified $139$ unique genes common to the $249$ CpG sites and $511$ probes from the gene expression data.
The selected methylation data $\bX_{179 \times 334}$ and gene expression data $\bY_{179 \times 511}$ were each normalized to have mean zero and variance one for each CpG site and gene.
We aim to identify a subset of methylation and genomic features that describe the overall association between CpG sites and gene expression. 

We applied the proposed sparse CCA via SELP on the methylation and gene expression data and compared with some existing sparse CCA methods. We randomly divided the data into three approximately equal-sized groups, with $2/3$rd of the data used as training set for selecting tuning parameters via 5- fold CV, and the $1/3$rd  as testing set for calculating the canonical correlation vectors and coefficients using the chosen tuning parameters. We repeated the process $20$ times. Comparisons of the methods are based on the number of variables selected from either set and canonical correlations obtained from the test data. 
Figure \ref{fig:realsettings1a} shows box plots of these measures from 20 repetitions.  As for the number of selected variables, SCCA and SCAD methods show much higher variability than CoLaR, PMD and the two SELP methods. It is especially noticeable that SELP-R selected five variables (medians at 4.89 and 5.11 respectively in either panel) almost all times, having very small variation. The SCCA, SCAD and CoLaR achieved much lower correlation coefficient. The PMD and SELP-I achieved highest correlation with small sets of variable, which shows their effectiveness as sparse methodologies. 

\subsection{Interpretation of Canonical Vectors}
\begin{figure}[htbp]
	%\begin{center}
	\begin{tabular}{lll}		
		\includegraphics[height = 1.3in]{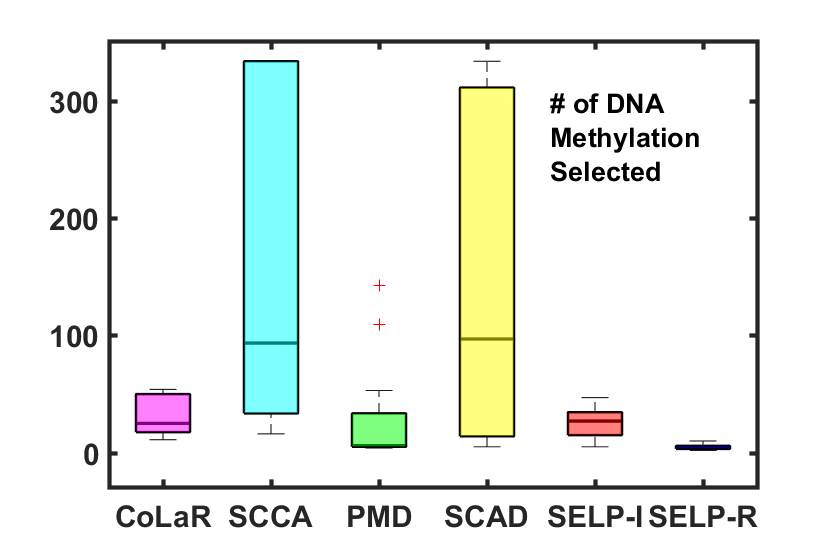}&\includegraphics[height = 1.3in]{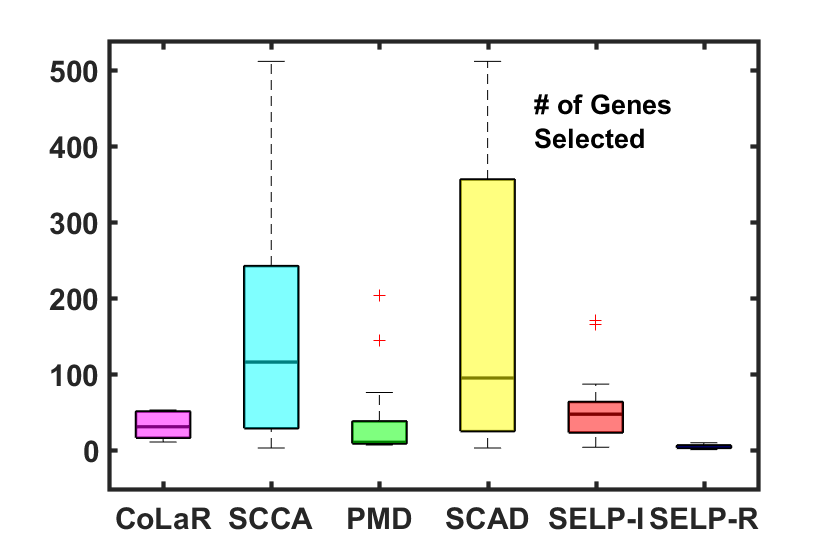}&\includegraphics[height = 1.3in]{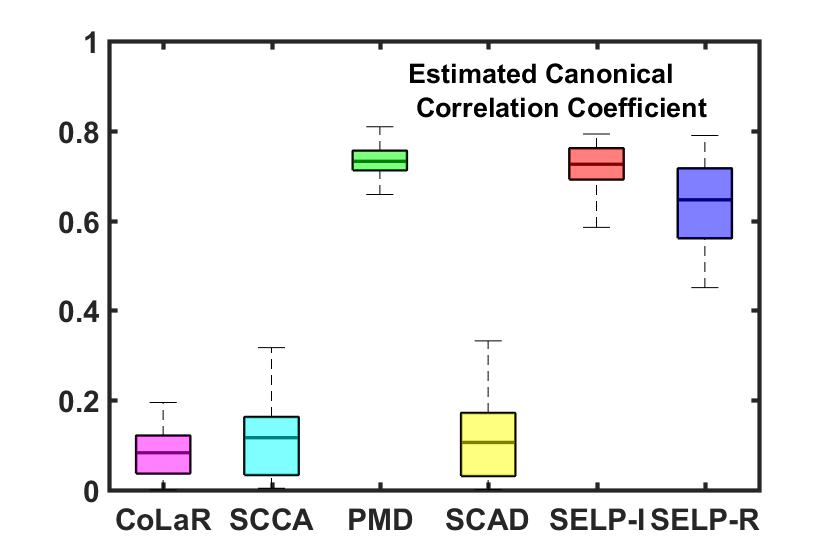}\\	
		~~~~~~~(a) Methylation  &~~~~~~~(b) Gene Expression &~~~~~~~ (c) Canonical Correlation\\
	\end{tabular}
	\vspace{0.7cm}
	\caption{Box plots of the number of Methylated DNA and genes that are selected,  and canonical correlation. The SELP methods have comparable or better estimated correlations and yet  select less genes and methylated DNA with less variation when compared to other sparse CCA methods. \label{fig:realsettings1a}}
	%\end{center}
	%\end{figure}
	
	%\begin{center}
	\begin{tabular}{ll}		
		\includegraphics[height = 2.0in]{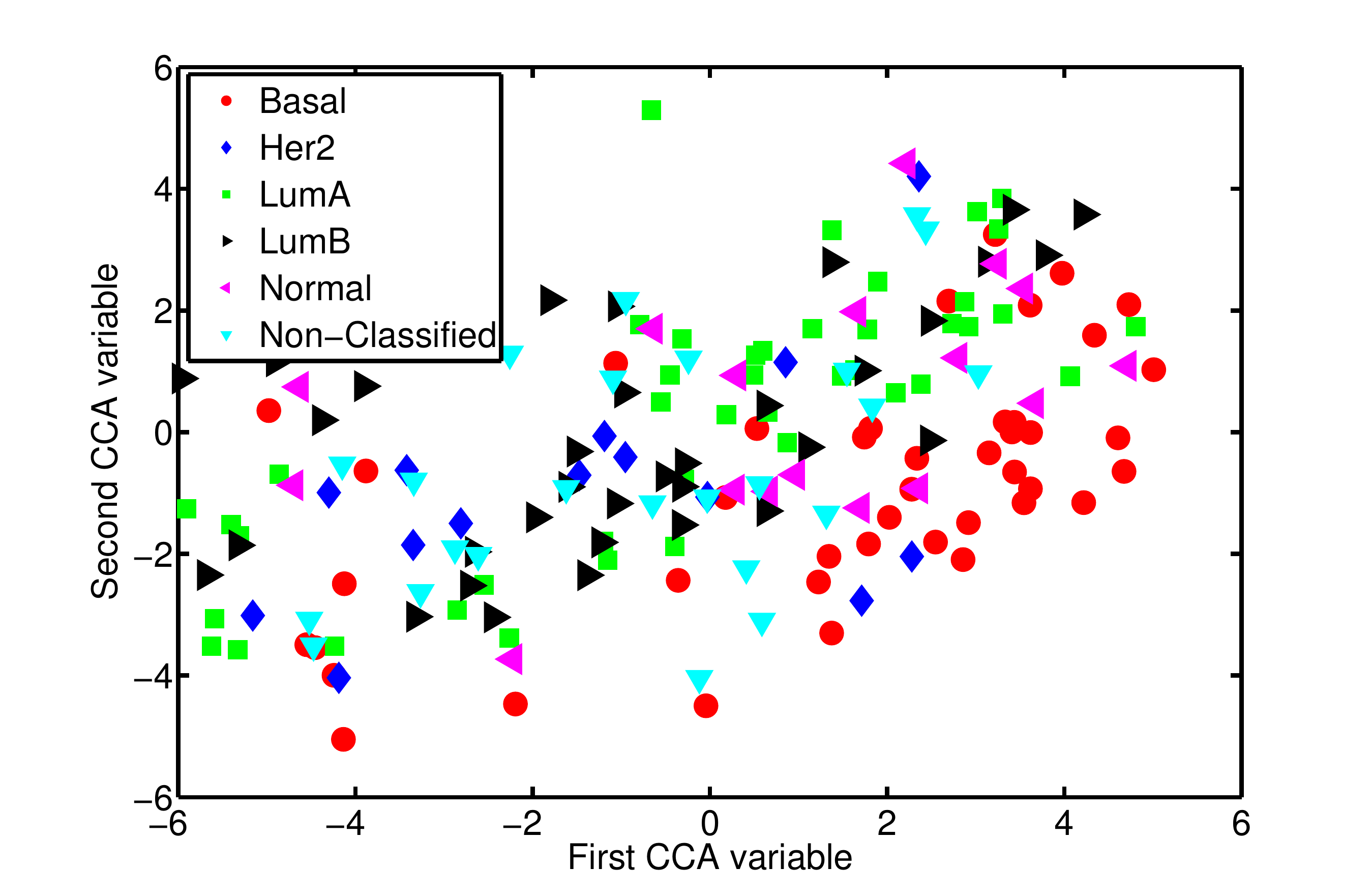}&\includegraphics[height = 2.0in]{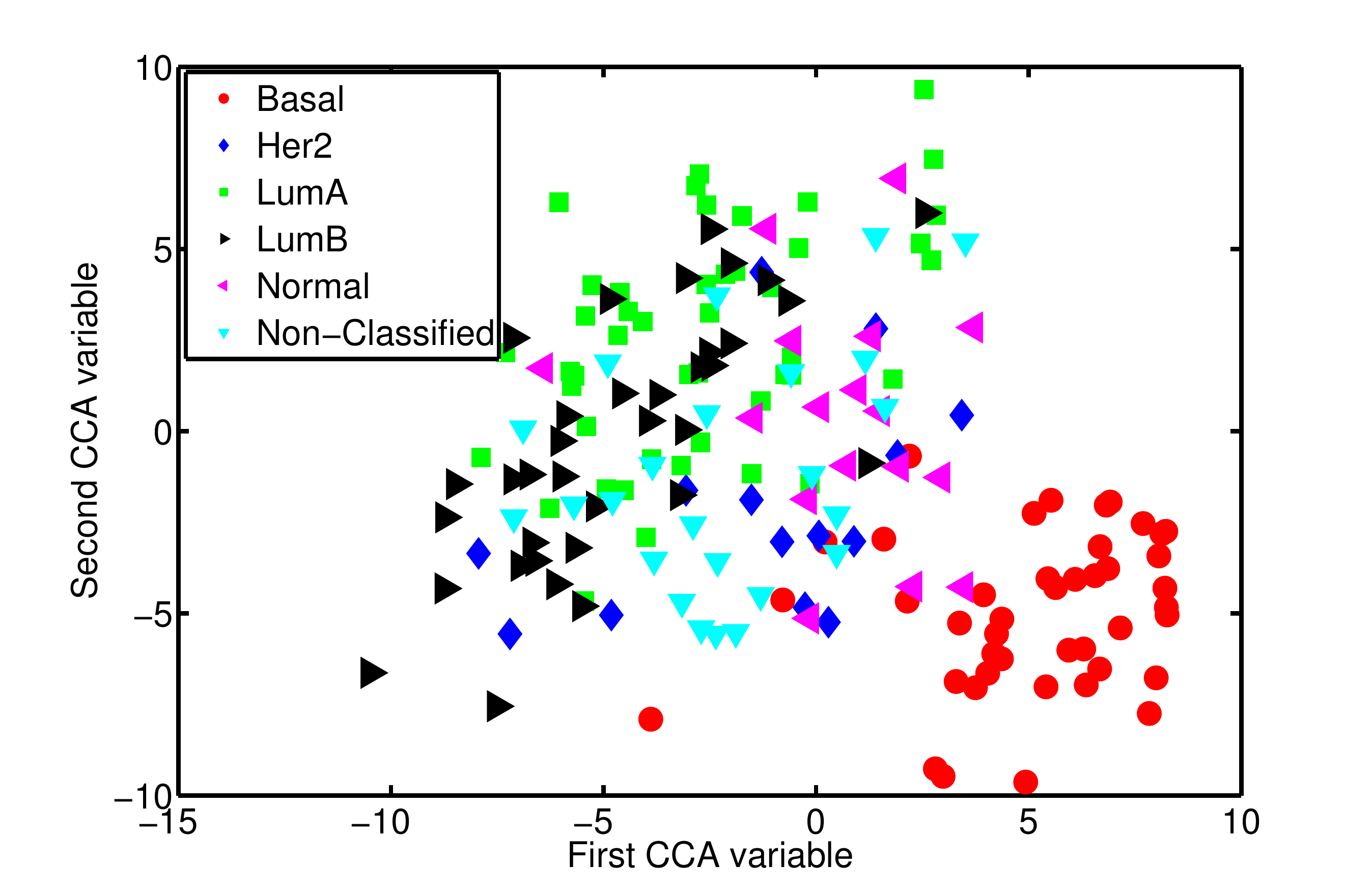}\\	
		~~~~~~~~~~~~~~~~~~ ~~~(a) Methylation  &~~~~~~~~~~(b) Gene Expression \\
	\end{tabular}
	\vspace{0.7cm}
	\caption{Projection scores of methylation and gene expression data onto the 2-dimensional canonical space.
		The basal-like breast cancer subtype is separated from other subtypes in gene expression data. The separation is less apparent in the methylation data.} \label{fig:realsettings1b}
	%\end{center}
	%\begin{figure}[htbp]
	%\begin{center}
	\begin{tabular}{ll}		
		\includegraphics[height = 2.2in]{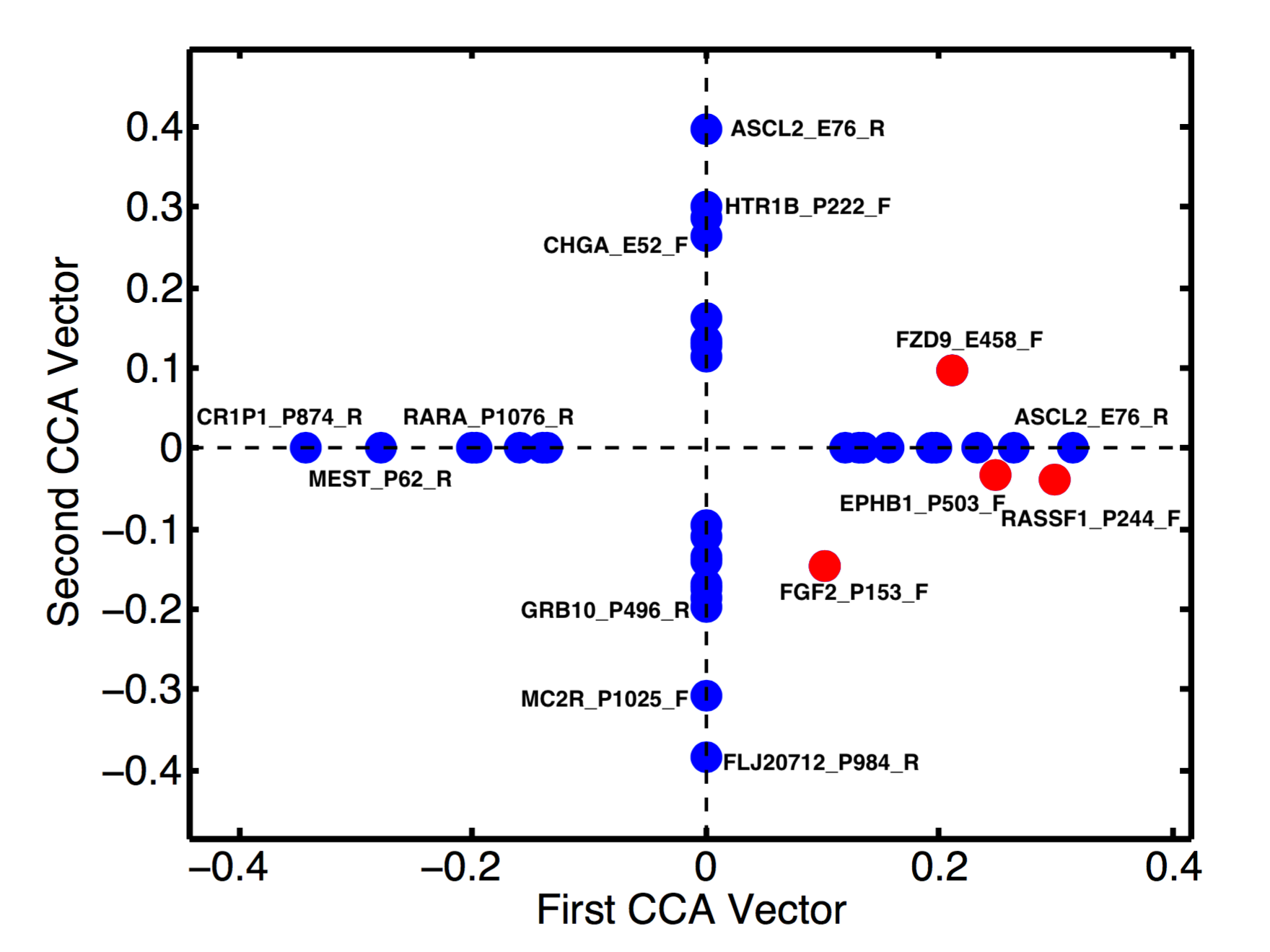}& \includegraphics[height = 2.2in]{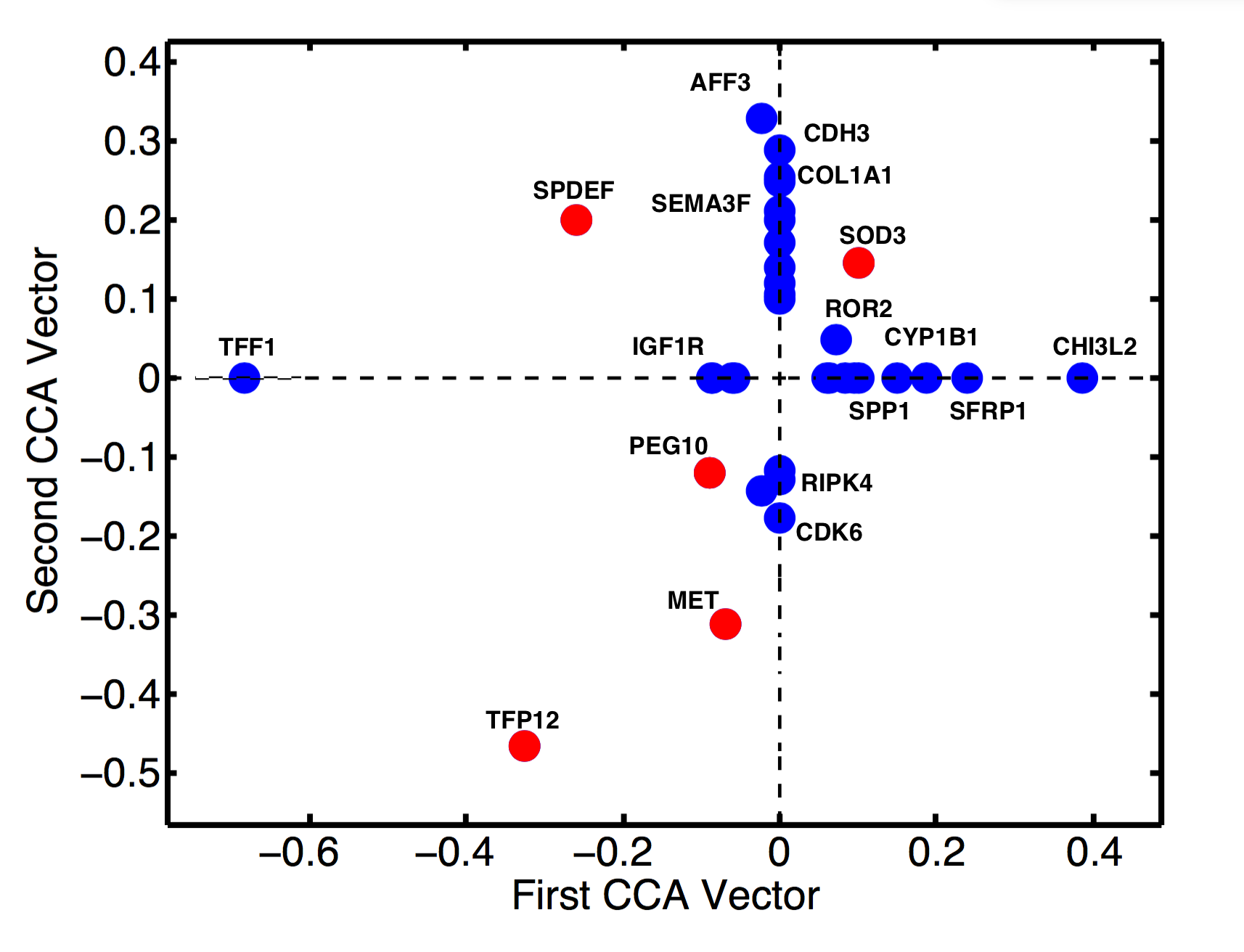}\\	
		~~~~~~~~~~~ ~~~(a) Methylation  &~~~~~~~~~~(b) Gene Expression \\
	\end{tabular}
	\vspace{0.7cm}
	\caption{DNA methylation and genes that are heavily weighted on the first and second canonical correlation vectors. Genes and DNA methylation labeled in blue have largest top twenty weights in the first and second CCA vectors. Genes  labeled in red are common in both axes.} \label{fig:realsettings1c}
	%\end{center}
\end{figure}

One can use the loadings of the canonical vector pairs to understand the contribution of each DNA methylation and gene on overall association between them. For this purpose we use the result of SELP-I, tune it via 5-fold CV and applied to the whole data. This approach identified $(45, 42)$ methylated DNA (corresponding to $(38, 42)$ unique genes) and $(38, 32)$ genes on the first and second canonical correlation vectors with correlations $0.41$ and $0.47$, respectively. We note the maximum canonical correlation coefficient is obtained on the second pair. Unlike the original CCA, this trend of nondecreasing canonical correlations is not unexpected in sparse CCA, due to the optimization criterion and regularization. This trend was also observed in \citep{WWZ:2008, Linetal:2013}. From the scatter plots in Figure \ref{fig:realsettings1b} of projection scores onto the canonical vectors, colored differently according to the breast cancer subtypes, we can see that the basal-like subtype is separated from the rest. Note that this visual separation is consistent with the existing cancer research works such as \citep{Nielsen:2004, Conway:2014}, where it has been found that patients with the basal-like subtype of breast cancer tend to have lower survival rates.

Figure  \ref{fig:realsettings1c} shows plots of the top twenty genes and DNA methylation with highest absolute loadings in the first and second canonical vectors. Blue dots are the genes or methylations heavily weighted on each axis, and red dots are common to both vectors. We conducted a further investigation of the specific functions and disease of the genes and DNA methylation identified  using ToppGene Suite \citep{chentoppgene:2009} for gene list enrichment analysis and DAVID (Database for Annotation, Visualization and Integrated Discovery) Bioinformatics Resources $6.7$ Functional Annotation Tool \citep{HSL:2009} for gene-gene ontology (GO) term enrichment analysis to identify the most relevant GO terms associated with the genes identified by the proposed method. Specifically, the genes that are selected for the first and second CCA vectors in the gene expression data are taken as input in the online tools to identify diseases and biological processes. Some of the significant biological processes enriched in our gene list include tissue development, epithelium development, cell proliferation, response to hormone, response to estrogen, negative regulation of response to stimulus, orderly cell division, cell motility  and negative regulation of cell differentiation. Also, some of the significantly enriched terms from our methylation gene list included secreted, cell motion, gland development, cell motility, regulation of cell development and cell proliferation and tumor suppressor. All these significant biological processes play essential roles in breast cancer progression or development.

We also investigated specific diseases associated with our gene lists. Fourteen genes from our gene expression data are identified to be associated with breast neoplasms (p-value $=1.335 \times 10^{-8}$)  which include breast tumors or  cancer, and these include the genes  subfamily B, polypeptide 1 (CYP1B1), secreted frizzled-related protein 1 (SFRP1), secreted phosphoprotein 1 (SPP1), tissue factor pathway inhibitor 2 (TFP12) and insulin-like growth factor 1 receptor (IGF1R).  Previous studies \citep{Klopockietal:2004} had demonstrated that SFRP1, which is heavily loaded on the first CCA vector (Figure \ref{fig:realsettings1c} (b)), is a putative inhibitor of Wnt signalling pathway \citep{Finchetal:1997} and that loss of SFRP1 protein expression is a common event in breast tumors that is associated with poor overall survival in patients with early breast cancer. Recently, \cite{Veecketal:2006} used mutation and methylation analysis to demonstrate that promoter hypermethylation is the predominant mechanism of SFRP1 gene silencing in human breast cancer and that SFRP1 gene inactivation in breast cancer is associated with unfavorable prognosis. In addition, the gene  Cystene-Rich Intestinal Protein 1 (CRIP1)  which is heavily loaded on the first CCA vector in the methylation data (Figure  \ref{fig:realsettings1c} (a)) is suggested to likely play a role as tumor cell proliferation suppressor in a breast cancer study \citep{Natalie:2013}.

It is worth mentioning that the proposed method successfully identified genes and methylated DNA that are known to be associated with breast tumorigenesis. Our results demonstrate that the proposed method generates biologically meaningful insights.

%\clearpage
\section{Discussion}\label{sec:discuss}
We have presented a method for individual and integrative analysis of high dimension, low sample size data. The major contributions of this paper are as follows: (1) to develop a general framework for obtaining sparse solution vectors for high dimension, low sample size problems and (2) to apply the framework to integrate different types of high dimension, low sample size data using canonical correlation analysis. We demonstrated the utility of the proposed method for joint analysis of methylation and gene expression data to identify methylated DNA that are highly associated with gene expression levels to understand better the etiology of breast cancer.

The proposed method capitalizes on a core idea in many multivariate statistical problems of extracting meaningful direction vectors spanning a lower dimensional subspace and their relationships with generalized eigenvalue problems. The solution from the traditional generalized eigenvalue problem is complicated by the dimension of the data; solution vectors  tend to yield results that do not induce sparsity, results that often times cannot be generalized and interpreted.  We proposed a methodology that approaches this difficult problem from a dimension-reduction perspective, using sparsity inducing penalty $l_{1}$ on the generalized eigenvalue vectors and constraining the difference of the generalized eigenvalue problem with $l_{\infty}$ norm, in order to look for lower dimensional subspace of few meaningful variables that can adequately summarize the data. 

It is worth mentioning that the SELP method can be applied to obtain sparse solution vectors in several multivariate statistical problems such as principal component analysis, linear discriminant analysis, multiple linear regression and multiple analysis of variance, to mention but a few. We leave it as future work to explore the performance of the SELP method in these problems.  Additionally, there is scope for generalizing the current  CCA formulation to more than two datasets. Another area for future work is to extend our method to nonlinear CCA to study nonlinear associations between methylation and gene expression data.

%\backmatter

%%%%%% include this section if you wish to acknowledge people,
%%%%%% grant support, etc.

%\section*{Acknowledgements}

%The work is supported by .\vspace*{-8pt}

%%%%%% include this section only if your manuscript refers to supplementary
%%%%%% materials -- see Instructions for Authors at 
%%%%%% http://www.tibs.org/biometrics

%\section*{Supplementary Materials}
%
%Web Appendix A referenced in Section~\ref{sec:theory} is available
%with this paper at the Biometrics website on Wiley Online Library.
%\vspace*{-8pt}
\section*{Software}
\label{software}
Software in the form of MATLAB code is available on request from the corresponding author (ssafo@emory.edu).\\

%\section*{Acknowledgments}

%The authors thank reviewers and editors for their useful comments.

\noindent{\it Conflict of Interest}: None declared.

%\clearpage
\bibliographystyle{apalike} 
\bibliography{refs}

\begin{thebibliography}{}

\bibitem[Cai and Liu, 2011]{CL:2011}
Cai, T. and Liu, W. (2011).
\newblock A direct estimation approach to sparse linear discriminant analysis.
\newblock {\em Journal of the American Statistical Association},
  106(496):1566--1577.

\bibitem[Candes and Tao, 2007]{Dantzig:2007}
Candes, E. and Tao, T. (2007).
\newblock {The Dantzig selector: Statistical estimation when $p$ is much larger
  than $n$}.
\newblock {\em The Annals of Statistics}, 35(6):2313--2351.

\bibitem[Chalise and Fridley, 2012]{CF:2012}
Chalise, P. and Fridley, B.~L. (2012).
\newblock Comparison of penalty functions for sparse canonical correlation
  analysis.
\newblock {\em Computational Statistics and Data Analysis}, 56:245--254.

\bibitem[Chen et~al., 2009]{chentoppgene:2009}
Chen, J., Bardes, E.~E., Aronow, B.~J., and Jegga, A.~G. (2009).
\newblock Toppgene suite for gene list enrichment analysis and candidate gene
  prioritization.
\newblock {\em Nucleic acids research}, 37(suppl 2):W305--W311.

\bibitem[Chen et~al., 2013]{Chenetal:2013}
Chen, M., Gao, C., Ren, Z., and Zhou, H.~H. (2013).
\newblock Sparse {CCA} via precision adjusted iterative thresholding.
\newblock \url{http://arxiv.org/abs/1311.6186}.

\bibitem[Conway et~al., 2014]{Conway:2014}
Conway, K., Edmiston, S.~N., May, R., Kuan, P.~F., Chu, H., Bryant, C., Tse,
  C.-K., Swift-Scanlan, T., Geradts, J., Troester, M.~A., and Millikan, R.~C.
  (2014).
\newblock Dna methylation profiling in the carolina breast cancer study defines
  cancer subclasses differing in clinicopathologic characteristics and
  survival.
\newblock {\em Breast Cancer Research : BCR}, 16(5):450.

\bibitem[CVX~Research, 2012]{cvx:2012}
CVX~Research, I. (2012).
\newblock {CVX}: Matlab software for disciplined convex programming, version
  2.0.
\newblock \url{http://cvxr.com/cvx}.

\bibitem[Dworkin et~al., 2009]{Dworkin:2009}
Dworkin, A.~M., Huang, T. H.-M., and Toland, A.~E. (2009).
\newblock Epigenetic alterations in the breast: Implications for breast cancer
  detection, prognosis and treatment.
\newblock {\em Seminars in cancer biology}, 19(3):165--171.

\bibitem[Elena et~al., 2009]{SCCA2:2009}
Elena, P., David, T., and Joseph, B. (2009).
\newblock Sparse canonical correlation analysis with application to genomic
  data integration.
\newblock \url{http://www.uhnres.utoronto.ca/labs/tritchler/scca.html}.

\bibitem[Fan and Li, 2001]{FL:2001}
Fan, J. and Li, R. (2001).
\newblock Variable selection via nonconcave penalized likelihood and its oracle
  properties.
\newblock {\em Journal of the American Statistical Association}, 96:1348--1360.

\bibitem[Finch et~al., 1997]{Finchetal:1997}
Finch, P.~W., He, X., Kelley, M.~J., Uren, A., Schaudies, R.~P., Popescu,
  N.~C., Rudikoff, S., Aaronson, S.~A., Varmus, H.~E., and Rubin, J.~S. (1997).
\newblock Purification and molecular cloning of a secreted, frizzled-related
  antagonist of wnt action.
\newblock {\em Proc Natl Acad Sci U S A}, 94(13):6770--6775.

\bibitem[Gao et~al., 2015a]{Chaoetal2:2015}
Gao, C., Ma, Z., Ren, Z., and Zhou, H.~H. (2015a).
\newblock Minimax estimation in sparse canonical correlation analysis.
\newblock {\em Annals of Statistics}, 43(5):2168--2197.

\bibitem[Gao et~al., 2015b]{Maetal:2015}
Gao, C., Ma, Z., and Zhou, H.~H. (2015b).
\newblock Sparse {CCA}: Adaptive estimation and computational barriers.
\newblock \url{http://arxiv.org/pdf/1409.8565.pdf}.

\bibitem[Hastie and Tibshirani, 2004]{HT:2004}
Hastie, T. and Tibshirani, R. (2004).
\newblock Efficient quadratic regularization for expression arrays.
\newblock {\em Biostatistics}, 5(3):329--340.

\bibitem[Holm et~al., 2010]{Holm:2010}
Holm, K., Hegardt, C., Staaf, J., Vallon-Christersson, J., Jonsson, G., Olsson,
  H., Borg, A., and Ringner, M. (2010).
\newblock Molecular subtypes of breast cancer are associated with
  characteristic dna methylation patterns.
\newblock {\em Breast Cancer Res}, 12(3):R36.

\bibitem[Hotelling, 1936]{Hotelling:1936}
Hotelling, H. (1936).
\newblock Relations between two sets of variables.
\newblock {\em Biometrika}, pages 312--377.

\bibitem[Huang et~al., 2009]{HSL:2009}
Huang, D.~W., Sherman, B.~T., and Lempicki, R.~A. (2009).
\newblock Systematic and integrative analysis of large gene lists using david
  bioinformatics resources.
\newblock {\em Nat Protocols}, 4(1):44--57.

\bibitem[Klopocki et~al., 2004]{Klopockietal:2004}
Klopocki, E., Kristiansen, G., Wild, P.~J., Klaman, I., Castanos-Velez, E.,
  Singer, G., Stohr, R., Simon, R., Sauter, G., Leibiger, H., Essers, L.,
  Weber, B., Hermann, K., Rosenthal, A., Hartmann, A., and Dahl, E. (2004).
\newblock Loss of sfrp1 is associated with breast cancer progression and poor
  prognosis in early stage tumors.
\newblock {\em Int J Oncol}, 25(3):641--649.

\bibitem[Lin et~al., 2013]{Linetal:2013}
Lin, D., Zhang, J., Li, J., Calhoun, V.~D., Deng, H.-W., and Wang, Y.-P.
  (2013).
\newblock Group sparse canonical correlation analysis for genomic data
  integration.
\newblock {\em BMC Bioinformatics}, 14:245.

\bibitem[Lock et~al., 2013]{Lock:2013}
Lock, E.~F., Hoadley, K.~A., Marron, J., and Nobel, A.~B. (2013).
\newblock Joint and individual variation explained (jive) for integrated
  analysis of multiple data types.
\newblock {\em The annals of applied statistics}, 7(1):523--542.

\bibitem[Ludyga et~al., 2013]{Natalie:2013}
Ludyga, N., Englert, S., Pflieger, K., Rauser, S., Braselmann, H., Walch, A.,
  Auer, G., H{\"o}fler, H., and Aubele, M. (2013).
\newblock The impact of cysteine-rich intestinal protein 1 (crip1) in human
  breast cancer.
\newblock {\em Molecular Cancer}, 12:28--28.

\bibitem[Mardia et~al., 2003]{MKB:2003}
Mardia, K.~V., Kent, J.~T., and Bibby, J.~M. (2003).
\newblock {\em Multivariate Analysis}.
\newblock Acadmeic Press.

\bibitem[Nielsen et~al., 2004]{Nielsen:2004}
Nielsen, T.~O., Hsu, F.~D., Jensen, K., Cheang, M., Karaca, G., Hu, Z.,
  Hernandez-Boussard, T., Livasy, C., Cowan, D., Dressler, L., Akslen, L.~A.,
  Ragaz, J., Gown, A.~M., Gilks, C.~B., van~de Rijn, M., and Perou, C.~M.
  (2004).
\newblock Immunohistochemical and clinical characterization of the basal-like
  subtype of invasive breast carcinoma.
\newblock {\em Clin Cancer Res}, 10(16):5367--5374.

\bibitem[Parkhomenko et~al., 2009]{PTB:2009}
Parkhomenko, E., Tritchler, D., and Beyene, J. (2009).
\newblock Sparse canonical correlation analysis with application to genomic
  data integration.
\newblock {\em Statistical Applications in Genetics and Molecular Biology}, 8.

\bibitem[Shen et~al., 2013]{shen:2013}
Shen, R., Wang, S., and Mo, Q. (2013).
\newblock Sparse integrative clustering of multiple omics data sets.
\newblock {\em Ann. Appl. Stat.}, 7(1):269--294.

\bibitem[Tibshirani, 1994]{Tibshirani:1994}
Tibshirani, R. (1994).
\newblock Regression shrinkage and selection via the lasso.
\newblock {\em Journal of the Royal Statistical Society, Series B},
  58:267--288.

\bibitem[Tibshirani et~al., 2005]{TSRZK:2005}
Tibshirani, R., Saunders, M., Rosset, S., Zhu, J., and Knight, K. (2005).
\newblock Sparsity and smoothness via the fused lasso.
\newblock {\em Journal of the Royal Statistical Society: Series B (Statistical
  Methodology)}, 67(1):91--108.

\bibitem[Veeck et~al., 2006]{Veecketal:2006}
Veeck, J., Niederacher, D., An, H., Klopocki, E., Wiesmann, F., Betz, B., Galm,
  O., Camara, O., Durst, M., Kristiansen, G., Huszka, C., Knuchel, R., and
  Dahl, E. (2006).
\newblock Aberrant methylation of the wnt antagonist sfrp1 in breast cancer is
  associated with unfavourable prognosis.
\newblock {\em Oncogene}, 25(24):3479--3488.

\bibitem[Waaijenborg et~al., 2008]{WWZ:2008}
Waaijenborg, S., de~Witt~Hamar, P. C.~V., and Zwinderman, A.~H. (2008).
\newblock Quantifying the association between gene expressions and dna-markers
  by penalized canonical correlation analysis.
\newblock {\em Statistical Applications in Genetics and Molecular Biology}, 7.

\bibitem[Witten et~al., 2013]{PMA:2013}
Witten, D., Tibshirani, R., Gross, S., and Narasimhan, B. (2013).
\newblock Penalized multivariate analysis.
\newblock \url{https://cran.r-project.org/web/packages/PMA/PMA.pdf}.

\bibitem[Witten et~al., 2009]{WTH:2009}
Witten, D., Tibshirani, R., and Hastie, T. (2009).
\newblock A penalized matrix decomposition, with applications to sparse
  prinicial components and canonical correlation analysis.
\newblock {\em Biostatistics}, 10(3):515--534.

\bibitem[Zou, 2006]{Zou:2006}
Zou, H. (2006).
\newblock {The Adaptive Lasso and Its Oracle Properties}.
\newblock {\em Journal of the American Statistical Association},
  101:{1418--1429}.

\bibitem[Zou and Hastie, 2005]{ZH:2005}
Zou, H. and Hastie, T. (2005).
\newblock Regularization and variable selection via the elastic net.
\newblock {\em Journal of the Royal Statistical Society, Series B},
  67:301--320.

\end{thebibliography}

\section*{Appendix}
\begin{proof}[Proof of Theorem 1]
	We first show that the true $\vv$ is feasible.
	Let $c' > c_2 +  c_1 s$.
	By the assumptions (\ref{eq:CriticalAssumption1}) and (\ref{eq:CriticalAssumption2}), the left-hand side of the constraint (\ref{eq:SELP-2}) is bounded as
	\begin{align}
	\| \Sv \tilde\vv - \Sv \vv \|_\infty   \nonumber
	& \le \| \Sv \tilde\vv - \Sc\vv \|_\infty + \| (\Sv - \Sc)\vv \|_\infty \nonumber \\
	&\le  \| \Sv \tilde\vv - \Sc\vv \|_\infty + \| (\Sv - \Sc)\|_{\rm{max}} \|\vv \|_1 \nonumber \\
	& \le (c_2 + c_1s ) \sqrt{\log p / n} \nonumber \\
	& \le c' \sqrt{\log p / n} \le \tau_n, \label{eq:feasible1}
	\end{align}
	where we have used the condition that $\vv$ is unit-length and $s$-sparse, which leads to $\|\vv\|_1 = \sum_{i=1}^p |\vv_i| \le s$. Thus, the true $\vv$ is in the feasible region, and
	\begin{align*}
	\|\hat\vv\|_1 \le \|\vv \|_1 \le s.
	\end{align*}
	
	For the rest of proof, we condition on the event that (\ref{eq:feasible1}) holds, which occurs with probability at least $1 - O(p^{-1})$.
	
	Next we show that $\|\hat\vv - \vv\|_\infty \le  4 M_0\tau_n$. In preparation, note that
	\begin{align*}
	\| \Sc \hat\vv - \Sc \vv \|_\infty
	&\le \| \Sv \hat\vv - \Sv\vv \|_\infty + \| (\Sc - \Sv)(\hat\vv - \vv)  \|_\infty \\
	&\le \| \Sv \hat\vv - \Sv\tilde\vv\|_\infty + \|\Sv\tilde\vv - \Sv\vv\|_\infty + \| (\Sc - \Sv)\|_{\rm{max}} \| \hat\vv - \vv \|_1 \\
	&\le \tau_n+ \tau_n + c_1\sqrt{\log p / n} (\| \hat\vv \|_1 +\| \vv \|_1 )\\
	&\le 2 \tau_n                              + 2 c_1 s \sqrt{\log p / n }  \\
	&\le 4 \tau_n.
	\end{align*}
	Then, since $\|\Sc\|_1 = \|\Sc\|_\infty \le M_0$ by the assumption and that $\Sc$ is symmetric,
	\begin{align} \label{eq:infinitenormbound}
	\| \hat\vv - \vv\|_\infty \le \|\Sc^{-1}\|_{\infty} \|\Sc (\hat\vv - \vv)\|_{\infty} \le 4 M_0  \tau_n.
	\end{align}
	
	Denote $T = \{ i : v_i \neq 0  \}$, and recall $s = |T|$. Let $t   = 4 M_0  \tau_n$.
	Define a thresholded version of $\hat\vv$ by $\hat{\vv}^t$ where
	$$ \hat{v}^t_i = \left\{
	\begin{array}{ll}
	\hat v_i, & \hbox{if $|\hat v_i| \ge 2t$;} \\
	0, & \hbox{if $|\hat v_i| < 2t$.}
	\end{array}
	\right.$$
	
	Since $\vv$ is feasible,
	\begin{align*}
	\|\vv\|_1 & \ge \| \hat\vv \|_1  \\
	& = \| \hat\vv^t \|_1 + \|\hat\vv^t - \hat\vv \|_1  \\
	& \ge \| \vv \|_1 - \| \hat\vv^t - \vv \|_1  + \|\hat\vv^t - \hat\vv \|_1,
	\end{align*}
	which in turn leads to
	\begin{align*}
	\|\hat\vv^t - \hat\vv \|_1 \le \| \hat\vv^t - \vv \|_1.
	\end{align*}
	This implies that
	\begin{align*}
	\| \hat\vv - \vv \|_1  \le \|\hat\vv^t - \hat\vv \|_1 + \| \hat\vv^t - \vv \|_1 \le 2 \| \hat\vv^t - \vv \|_1.
	\end{align*}
	Therefore, an upper bound of $\| \hat\vv - \vv \|_1 $ can be obtained by a bound for $\| \hat\vv^t - \vv \|_1$.
	
	Note that for $i \notin T$, $ v_i = 0$,  and together with the maximum norm bound (\ref{eq:infinitenormbound}), we get $|\hat v_i| \le t$, which in turn leads that $ \hat v_i^t  = 0$. Thus, for all $i \notin T$, $\hat v^t_i -  v_i = 0$. On the other hand, if $i \in T$ and $\hat v^t_i = 0$, then again by (\ref{eq:infinitenormbound}), $| v_i| \le 3t$.
	Thus we can write
	\begin{align}
	\|  \hat\vv^t - \vv \|_2^2 & = \sum_{i \in T} (\hat v_i^t -  v_i)^2   \nonumber \\
	& = \sum_{i \in T} (\hat v_i^t -  v_i)^2 1_{\{\hat v_i^t = 0\}}
	+ \sum_{i \in T} (\hat v_i^t - v_i)^2 1_{\{\hat v_i^t \neq 0\}}  \nonumber \\
	& = \sum_{i \in T} v_i^2 1_{\{ \hat v_i^t = 0 \}}
	+ \sum_{i \in T} (\hat v_i - v_i)^2 1_{\{ \hat v_i^t \neq 0\} }  \nonumber  \\
	& \le \sum_{i \in T}  v_i^2 1_{\{ | v_i| \le 3t \}}
	+ \sum_{i \in T} t^2 1_{\{ \hat v_i^t \neq 0\}}    \nonumber  \\
	& \le  10 s t^2   \label{eq:1}
	\end{align}
	where in the first inequality we again used  (\ref{eq:infinitenormbound}). Then,
	\begin{align*}
	\| \hat\vv - \vv \|_1 &\le  2 \| \hat\vv^t - \vv \|_1  \\
	& =   2\sum_{i \in T} | \hat v^t_i - v_i |   \\
	&\le 2 \sqrt{s} (\sum_{i \in T} (\hat v^t_i - v_i)^2 )^{1/2}  \\
	&\le 2 \sqrt{10} st   =  8 \sqrt{10} s M_0  \tau_n.
	\end{align*}
	This shows (\ref{eq:thmL1}).
	
	For (\ref{eq:thmL2}), we first evaluate an upper bound on $\| \hat\vv - \hat\vv^t  \|_2$.
	Note that for $i \in T$, if $\hat v^t_i = 0$, then $ | \hat v_i|  \le 2t$, and if $\hat v^t_i \neq 0$, then $(\hat v_i - \hat v^t_i)^2 = 0 $.
	On the other hand, for $i \notin T$, $(\hat v_i - \hat v^t_i)^2 = \hat v_i^2$ and $\hat v_i \le t$. Moreover, we get
	\begin{align*}
	\sum_{i \notin T} | \hat v_i | &= \sum_{i \notin T} | \hat v_i -  v_i | \\
	& \le \sum_{i} | \hat v_i -  v_i | = \| \hat\vv - \vv \|_1 \\
	& \le 2 \sqrt{10} st.
	\end{align*}
	It turns out
	\begin{align}
	\| \hat\vv - \hat\vv^t  \|^2_2 & =
	\sum_{i\in T} (\hat v_i - \hat v^t_i )^2
	+ \sum_{i \notin T} (\hat v_i - \hat v^t_i )^2   \nonumber \\
	& \le \sum_{i\in T} (\hat v_i - \hat v^t_i )^2
	+ \sum_{i \notin T} \hat v_i^2   \nonumber \\
	& \le 4st^2  + \max_{i \notin T } |\hat v_i | \sum_{i \notin T}        |\hat v_i|  \nonumber \\
	& \le (4   +  2 \sqrt{10} ) st^2. \label{eq:2}
	\end{align}
	Therefore, (\ref{eq:thmL2}) is obtained by combining (\ref{eq:2}) and (\ref{eq:1}) as follows:
	\begin{align*}
	\| \hat\vv - \vv  \|_2 &\le \| \hat\vv - \hat\vv^t  \|_2 + \| \hat\vv^t - \vv  \|_2 \\
	&\le   ( (4   +  2 \sqrt{10} )^{1/2}   + \sqrt{10}) \sqrt{s}t.
	\end{align*}
\end{proof}

%\label{lastpage}

\end{document}